\documentclass[fleqn,usenatbib]{mnras}
\usepackage{amsmath}	
\usepackage{txfonts}
\usepackage[T1]{fontenc}
\usepackage{ae,aecompl}

\usepackage{graphicx}	

\newcommand\un[1]{{\,\rm #1}}
\newcommand\E[1]{\times10^{#1}}
\newcommand\rs[1]{_\mathrm{#1}}

\title[MHD simulations of young SNRs]{Magneto-hydrodynamic simulations of young supernova remnants {and their energy-conversion phase}}

\author[O.Petruk, T.Kuzyo, S.Orlando, M.Pohl, R.Brose]{O. Petruk$^{1,2}$, T. Kuzyo$^{1}$, S. Orlando$^{3}$, M.Pohl$^{4}$, R.Brose$^{4,5}$\\
$^{1}$Institute for Applied Problems in Mechanics and Mathematics, Naukova St 3-b, 79060 Lviv, Ukraine\\
{$^{2}$Astronomical Observatory of Taras Shevchenko National University of Kyiv, Observatorna St 3, 04053 Kyiv, Ukraine}\\
$^{3}$INAF -- Osservatorio Astronomico di Palermo, Piazza del Parlamento 1, I-90134 Palermo, Italy\\
$^{4}$DESY,  Platanenallee 6, 15738 Zeuthen, Germany\\
{$^{5}$Dublin Institute for Advanced Studies, Astronomy \& Astrophysics Section, 31 Fitzwilliam Place, D02 XF86 Dublin 2, Ireland}}

\date{Accepted XXX. Received YYY; in original form ZZZ}

\pubyear{2020}

\begin{document}
\label{firstpage}
\pagerange{\pageref{firstpage}--\pageref{lastpage}}
\maketitle

\begin{abstract}
    Supernova remnants (SNRs) can be rich sources of information on the parent SN explosion. Thus investigating the transition from the phase of SN to that of SNR can be crucial to link these two phases of evolution. Here we aim to study the early development of SNR in more details, paying the major attention to the transition from the early-expansion stage to the Sedov stage and the role played by magnetic field in this transition. To this end, spherical magneto-hydrodynamic simulations of SNRs have been performed to study the evolution of magnetic field in young SNRs and explore a sequence of the SNR evolutionary stages in the pre-radiative epoch. Remnants of three supernova types are considered, namely, SNIa, SNIc and SNIIP, that covers a wide space of parameters relevant for SNRs. Changes in global characteristics and development of spatial distributions are analysed. It is shown that the radial component of magnetic field rapidly drops downstream of the forward shock. Therefore, the radially-aligned polarization patterns observed in few young SNRs cannot be reproduced in the one-dimensional MHD simulations. The period SNR takes for the transition from the earliest ejecta-driven phase to the Sedov phase is long enough, with its distinctive physical features, headed by the energy conversion from mostly kinetic one to a fixed ratio between the thermal and kinetic components. This transition worth to be distinguished as a phase in SNR evolutionary scheme. The updated sequence of stages in SNR evolution could be the free expansion (of gas) -- energy-conversion -- Sedov-Taylor -- post-adiabatic -- radiative.
\end{abstract}


\section{Introduction}

There are two kind of experiments which may be used to decode information from supernova remnants (SNRs) about their progenitor star, properties of a supernova (SN) explosion and physics of shocks as well as the structure of an ambient medium, namely, observations and simulations. There are just few young -- with the age, say, less than 1000 years -- SNRs known. Moreover, there are big time gaps between the youngest SNRs that prevent us from reconstructing the smooth evolution of SNRs from their observations. Numerical experiments are therefore vital for our knowledge about transformation of SN to SNR and in gaining physical insight of the phenomena that occurred during a SN and of the nature of the progenitor star (e.g. \citealt{2014ApJ...791...97L,
2015ApJ...803..101P, 2015ApJ...810..168O, 2016ApJ...822...22O, 2020A&A...636A..22O, 2021A&A...645A..66O, 2021MNRAS.502.3264G} in the case of core-collapse (CC) SNe, and \citealt{2019ApJ...877..136F, 2021ApJ...906...93F} in the case of type Ia SNe). 

There are two main SN explosion mechanisms: thermonuclear explosion of a white dwarf in a tight binary system (type Ia) and core collapse of a massive star (type Ib/c and II).

\citet{1982ApJ...258..790C,1985Ap&SS.112..225N} have shown that the early stages of SNR evolution with the power-law density ($\rho\rs{ej}\propto r^{-n}$, $n>5$) ejecta expanding into stationary ambient medium with power-law density distribution ($\rho\rs{o}\propto r^{-s}$) may be described by the 'self-similar driven wave' solutions of hydrodynamics (HD) equations. 
The model considers the SN ejecta expansion into the circumstellar medium (CSM) and results in formation of a double-shock structure. 
The structure consists of two gas shells,  
shocked ISM and shocked stellar ejecta, separated by a contact discontinuity (CD) and bounded by the forward (FS) and reverse (RS) shocks respectively.
The self-similar solution can be used both for type I explosion in a uniform CSM and type II SN events evolving in the wind-blown bubble CSM. 
\citet{1984ApJ...281..682H} demonstrates the free expansion of CD for $n<5$. 

\citet{1982ApJ...259..302C} has introduced an approximate model for the interaction of the type II SN shock  with the power-law CSM. It assumes the zero distance between RS and FS and gives an insight into the most important physical features determining the overall shock dynamics at this epoch, as well as  derives an equation for the shock motion. 

The features of the early interaction of an exploded massive star with its circumstellar wind bubble can be described by another semi-analytical solutions which assumes the power-law supernova ejecta with a flat core \citep{1989ApJ...344..332C}. 
After the interaction with the bubble shell the post-shock flow gradually evolves towards the \citet{1982ApJ...258..790C} solutions.

\citet{1998ApJ...497..807D} have demonstrated that the exponential ejecta profile describes SNIa evolution  more accurately compared to the power-law profile.
The authors analysed differences in the flow evolution between the power-law and exponential ejecta models.

The particle acceleration by the young remnants of both SN types has been investigated by
\citet[][SNIa]{2012APh....35..300T}, \citet[][CC SN]{2013A&A...552A.102T}.
It is shown that the structure of the post-shock flow strongly affects the acceleration of CRs both at the forward and at the reverse shock. 
The authors have demonstrated how the hydrodynamics of the post-shock flow and features of interaction of SNR with ISM determine the accelerated particle spectrum and therefore the emission intensity and shape the emission spectra, with significant differences between the type-Ia and the core-collapse SNe.

The evolution of magnetic field in SNR is underexplored though its role on e.g. shaping SNRs is under consideration for a long time \citep[e.g.][]{1962MNRAS.124..179V}. 
{Fully \textit{magneto-}hydrodynamic (MHD) simulations of SNRs, i.e., by joint solution of MHD equations, are {not common}  \citep[e.g.][]{2007A&A...470..927O,2012ApJ...749..156O,2015A&A...579A..35Y,2017MNRAS.466.4851V,2019A&A...622A..73O,2020MNRAS.494.1531M}}.
More popular approach consists in a simplified prescription of magnetic field in Sedov SNRs. It is used in cases where one need to calculate SNR synchrotron emission \citep[e.g.][]{1974ApJ...188..501C,1998ApJ...493..375R,2017MNRAS.470.1156P}. It is based on the equations for magnetic field \citep[see references in Sect.2.2 in][]{2016MNRAS.456.2343P} applied to the spatial SNR structure derived from the HD solutions or simulations. 
Consideration of MF on the pure HD structure is justified for the evolved non-radiative SNRs (Sedov stage) because the magnetic energy density for typical SNR conditions is considerably lower than the thermal energy density, so the effect from magnetic field on the  structure and dynamics in such SNRs may be neglected \citep[e.g.][]{1972ARA&A..10..129W}. Instead, MHD simulations have shown that magnetic field and its orientation is crucial in evolution of the post-adiabatic shocks \citep{2016MNRAS.456.2343P,2018MNRAS.479.4253P}; in particular, by limiting the compression of cooling plasma, as predicted by \citet{1972ARA&A..10..129W}.

The goal of the present paper is to study the early evolution of SNR in more details, paying the major attention 
(i) to the transition from the early-expansion stage to the Sedov stage as well as 
(ii) to the role and behaviour of magnetic field. 
Our model and numerical setups are described in Sect.~\ref{youngsnr:setup}. Sect.~\ref{youngsnr:shock} is devoted to the shock kinetics and properties of SNR in general. Internal structure, including magnetic field, is considered in Sect.~\ref{youngsnr:interior}. Some important points are discussed in Sect.~\ref{youngsnr:discuss}. Sect.~\ref{youngsnr:conclusions} concludes. Animations relevant for our study are presented in Appendix.

\section{Problem Setup}
\label{youngsnr:setup}

We consider three supernova event types representing both physical explosion mechanisms: 
a thermonuclear explosion (Ia) and core-collapse events (Ic and IIP). 
Since the core-collapse (CC) SNe span a wide range of progenitor star and circumstellar medium (CSM) parameters, we select only two  scenarios, while parameters of others' SNe fall between these two. 
Conversely, thermonuclear SNe being standard candles in cosmology demonstrate rather narrow space of parameters. 

For each of the SN types we run spherically-symmetric numerical magneto-hydrodynamic (MHD) simulations starting from several days (namely, $t\rs{o}=0.01\un{yr}$) since the explosion up to a well-established radiative stage. 
We use the initial shock velocity of $V\rs{o}=30 000 \un{km/s}$ for all our runs, which at $t\rs{o} = 0.01 \un{yr}$ corresponds to the radius of a boundary between ejecta and ISM of $R\rs{o}=V\rs{o}t\rs{o}= 3\times10^{-4} \un{pc}$. 

\subsection{SNIa explosion}

Type Ia SNRs result from white dwarf explosions in tight binary systems.
Typically the system evolves in more or less uniform CSM and ISM; in our model therefore the shock propagates into a medium with uniform distribution of density and magnetic field (MF). Given the lack of information on the configuration of the magnetic field in the CSM around these systems and (at later times) in the ISM, we decided to consider the simplest configuration of a uniform (on the scale of the simulated system) magnetic field. 
A constant B-field is a reasonable assumption for the ISM around a type-Ia SN and around the wind bubble of a core-collapse SN. In the shocked wind of the progenitor, the tangential field tends to slowly increase with radius, and radial field falls off, but stellar evolution and shock reflections lead to turbulent transport that likely flattens the distributions. The choice of a uniform magnetic field is further  justified by the fact that the magnetic field is not expected to affect the overall evolution of the large scale structure of the remnant given the high plasma $\beta$ characterizing the plasma. For our purposes it is enough to introduce a seed magnetic field which allows us to study its effects on the shock structure during the transition from the phase of SN to that of SNR. We will consider both limiting MF orientations, radial and tangential; any other orientation is just a combination of these two.

Initially, the supernova ejecta has exponential density profile \citep{1998ApJ...497..807D}:
\begin{equation}
    \rho\rs{ej} = A \exp(-v / v\rs{e}) t^{-3}
    \label{eq_DC_ejecta_profile}
\end{equation}
where $t$ is the time since the explosion; constants $A$ and $v\rs{e}$ are obtained by integration 
over the ejecta density to find the total ejecta mass $M\rs{ej}$ and kinetic energy $E\rs{o}$
(almost 99.9\% of SN explosion energy is deposited in the kinetic component). 
Initially, the ejecta velocity is a linear function of the radial distance: $v(r) = r / t$ corresponding to the homologous expansion of the stellar interior.

The solution of the system
\begin{align}
    M\rs{ej} & = 4\pi \int_{0}^{\infty} \rho\rs{ej} r^2dr, \\
    E\rs{o}  & = 4\pi \int_{0}^{\infty} \frac{ \rho\rs{ej} v^2}{2} r^2dr
\end{align}
gives expressions for $v_e$ and $A$:
\begin{align}
    v\rs{e} & = \sqrt{\frac{E\rs{o}}{6M\rs{ej}}}, \\
    A & = \frac{6^{3/2}}{8\pi}\frac{M\rs{ej}^{5/2}}{E\rs{o}^{3/2}}.
\end{align}

We use parameters of typical white dwarf undergoing the thermonuclear explosion with the mass of $1.4 \un{M_\odot}$ 
and the energy of $10^{51} \un{erg}$. We set the initial ejecta temperature everywhere to $10^{7}\un{K}$.
The hydrogen number density of the uniform ISM is taken $n\rs{o} = 1\un{cm^{-3}}$.

\subsection{Core-Collapse explosion}

Following \citet{2013A&A...552A.102T}, we consider two boundary scenarios for the core-collapse SN events, namely
type-IIP (RSG progenitor) and type-Ic (WR progenitor). In both cases one can employ the same approach -- 
 a power-law ejecta density profile with a flat core as described by \citet{1989ApJ...344..332C}:
\begin{equation}
    \rho\rs{ej} =
    \begin{cases}
        Ft^{-3},& v\leq v\rs{t}\\
        Ft^{-3}(v/v\rs{t})^{-n} , & v > v\rs{t}
    \end{cases}
    \label{eq_CL_ejecta_profile}
\end{equation}
with
\begin{equation}
    F = \dfrac{1}{4\pi n} \dfrac{\left[3 (n-3) M\rs{ej} \right]^{5/2}}{\left[10 (n-5) E\rs{o}\right]^{3/2}}
\end{equation}
and transition velocity 
\begin{equation}
    v\rs{t} = \sqrt{\dfrac{10 (n-5) E\rs{o}}{3 (n-3) M\rs{ej}}}.
\end{equation}
The power-law index $n$ represents the ejecta profile steepness. 
The initial velocity profile for CC SNe {reflects homologous expansion ($v(r) = r / t$)} as in the case of type-Ia model.

Type-IIP progenitor stars typically have masses $8-20 \un{M\rs{\odot}}$ and ejecta steepness close to $n = 11$ \citep{MM99}. Stars producing SNIc posses larger masses $M\rs{ej} > 35\un{M\rs{\odot}}$ but {shallower} density profiles with $n = 9$ \citep{2005ApJ...619..839C}.
The mass of the exploding ejecta is less than the overall star mass   \citep[see Fig. 12 in][]{2017hsn..book.1879H}, considering that the star constantly loses its mass by the wind during the evolution. Namely, the pre-supernova ejecta doesn't exceed $15\un{M\rs{\odot}}$ for Type-Ic SNe and is less than $8\un{M\rs{\odot}}$ for type-IIP.

We adopt the flat innermost ejecta profile \citep{1999ApJS..120..299T, 2005ApJ...619..839C} in order to avoid singularity and thus have a finite density at the center (the core radius, as from equation~(\ref{eq_CL_ejecta_profile}), is $r\rs{t} \approx 0.01 r\rs{o}$). Other possibility for the inner ejecta core, the power-law distribution (like the outer part but with more flat $n < 3$ power-law index) is also considered \citep{MM99,2005ApJ...619..839C}. The choice does not affect the physics of FS and CD evolution; though, it could modify RS velocity when RS approaches the explosion center; the overall effect on the SNR dynamics is negligible.

The ejecta temperature exponentially decreases from $10^7\un{K}$ in the explosion center to the ISM value of $10^4\un{K}$ at the ejecta radius.

Unlike type Ia SN explosions, the CSM around the core-collapse supernovae is shaped by the stellar wind bubble. 
With constant wind parameters, the bubble density is defined by a commonly used expression
\begin{equation}
    \rho\rs{CS} = \dfrac{\dot{M}}{4\pi r^2 v\rs{w}} = D r^{-2}
    \label{eq_stellar_wind_profile}
\end{equation}
where $\dot{M}$ is progenitor's mass loss rate and $v\rs{w}$ is a wind speed.
In the case of RSG, typical values of $v\rs{w}$ are about $10 \un{km s^{-1}}$ and mass loss rate
is $5\cdot10^{-5} \un{M\rs{\odot}\ yr^{-1}}$ \citep{2005ApJ...630..892D}. 
Bubbles around WR stars, on the other hand, have much larger wind speed values, of order $10^3 \un{km\ s^{-1}}$, 
but lower mass loss rates $10^{-5} \un{M\rs{\odot}\ yr^{-1}}$ \citep{2005ApJ...619..839C}. Large wind speed results
in a large fraction of stellar material of the progenitor star being lost by the wind \citep{2007ApJ...667..226D}. We take the same parameter $D$ as {used for Fig.~1 in the paper} \citet{2013A&A...552A.102T}, {cf. Table \ref{table_model_param}}.

\begin{figure}
  \centering 
  \includegraphics[width=\columnwidth, trim={12pt 12pt 12pt 10pt},clip]{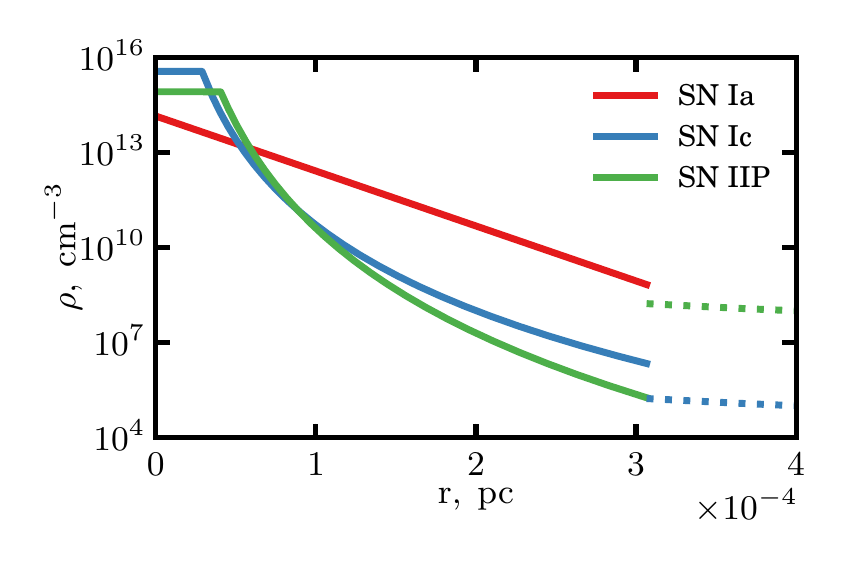}
  \caption{
    Radial density profiles for supernova ejecta (solid) and CSM bubbles (dotted) in the vicinity of the supernova at $t\rs{o}=0.01\un{yr}$ for our three explosion models.  
    SNR after type-Ia event evolves in a uniform ISM with $n\rs{o} = 1\un{cm^{-3}}$ (not shown here). 
  }
  \label{fig_ejecta_bubbles}
\end{figure}

Fig. \ref{fig_ejecta_bubbles} shows the stellar ejecta density profiles of the three described models with their CSM shaped by the stellar wind (for CC explosions). 
Based on mass loss rates and wind velocities one can clearly see a significant difference in CSM density of several orders of magnitude between the two CC SN cases.\footnote{Under parameters considered, the SNIIP wind profile has higher density than the ejecta at $R\rs{o}$; apparently, this only affects the propagation of the reverse shock up to $0.5\un{yrs}$ (Fig. \ref{fig_SN_all_r_ratio_shock_wave}b) that is unessential for our purposes.}

We assume that there is a smooth transition from CSM wind ($s=2$) to uniform ISM ($s=0$) at some distance, so that the ambient medium density is $\rho\rs{o}(r) = \rho\rs{CS}(r) + \rho\rs{ISM}(r)$. This transition allows us to avoid irregularities of the picture of SNR evolution by rather complex interactions of the forward shock with a sharp CSM bubble's edge. MF model is the simplest possible; it is same as for the SNIa model.

\subsection{MHD Code and Numerical Setup}

We use the PLUTO Code \citep{2007ApJS..170..228M,2012ApJS..198....7M} to solve the system of MHD equations for SNR evolution.
The computational setup is composed of \verb!RK3! time stepping and \verb!HLL! Riemann solver for spatial integration.
For all simulation runs we use spherically symmetrical (1D) MHD models on a static uniform grid with 200 000 zones. Such a high resolution which is only possible in 1-D simulations allows us to ignore the diffusivity of the solver and obtain the very accurate profiles including the shock jumps.
The boundary conditions at the end of the computational domain provide zero gradient while reflective conditions (rigid wall) are applied at the beginning of the domain. 

The simulations were carried starting from 0.01 yrs to about 500 000 yrs, long enough for SNRs to enter the radiative evolution stage.
Because the entire SNR evolution spans several orders of magnitude on both spatial and temporal scales,
one needs an approach which both provides decent spatial resolution near the shock front and preserves the post-shock flow features.
For that purpose we have employed a re-gridding technique as following:
every time the shock front approaches to the edge of the computational domain 
(shock radius is more than 90\% of the domain size),
we switch to a larger computational grid with the same number of zones 
but a larger physical extent and interpolate data from the smaller into the larger grid.
The physical extent of the initial grid is 0.2 pc and re-gridding factor is 5, so during the simulations the grid 
extent changes from 0.2 pc to 1 pc, then to 5 pc, 25 pc and 125 pc. The re-grid time depends on the SNR model and is summarized in the Table~\ref{table_model_param}. Considering the very large resolution of our simulations, mass, momentum and energy components are preserved with a very high accuracy during the re-gridding (this will be visible later at Figs.~\ref{fig_SN_all_mass_momentum} and \ref{fig_SN_all_energy_components}). The ejecta in the initial grid spans about 600 zones. 

In  our models, the adiabatic index $\gamma=5/3$ and the temperature of the ambient medium is set to $10\,000\un{K}$, which corresponds to the sound speed about 15 km/s ($c\rs{s} = \sqrt{\gamma R T / \mu}$) for the cosmic abundances ($\mu = 13/21$). 

Ambient magnetic field is uniform and has the strength $B\rs{o} = 10\un{\mu G}$. We consider two different orientations of $\mathbf{B}\rs{o}$: the shock is either parallel or perpendicular depending on the orientation of $\mathbf{B}\rs{o}$ and the shock normal. 
Generally, the tangential orientation of magnetic field (perpendicular shock) has the largest impact on the post-adiabatic flows. 
A seed magnetic field is needed also in the initial remnant interior. 
In our simulations, considering that the configuration of MF in the stellar interior is poorly known and the simulations are just 1-D, MF in the ejecta is taken nonuniform but simple, in a way which provide the local magnetic pressure to be $10\%$ of the local thermal pressure. 
The jump in the magnetic field strength at the edge $R\rs{o}$ from ejecta to the ambient medium is a few orders of magnitude for all our models.

Table \ref{table_model_param} summarizes details of our numerical  simulations for different SN explosion events.

\begin{table}
	\centering
	\caption{SNR model parameters and details of the runs.}
	\label{tab:example_table}
	\begin{tabular}{l|rrr} 
		\hline
		Model parameters & SN Ia & SN Ic & SN IIP\\
		\hline
		Explosion energy $E\rs{o}$, $10^{51}\un{erg}$ & 1 & 1 & 1\\
		Ejecta mass $M\rs{ej}$, $M_\odot$ & 1.4 & 14 & 8\\
		Ejecta steepness $n$ & - & 9 & 11\\
		CSM density steepness $s$  & 0 & 2;0 & 2;0\\
		CS bubble parameter $D$, {$\un{pc}^{2}/\un{cm}^{3}$} & - & 0.016 & 16\\
		ISM H density $n\rs{ISM}$, $\un{cm^{-3}}$ & 1 & 1 & 1\\ 
		MF strength $B\rs{o}$, $\mathrm{\mu G}$ & 10 & 10 & 10\\ 
		Initial shock velocity $V\rs{o}$, $\un{km/s}$ & $30 000$ & $30 000$& $30 000$\\
		Initial time $t\rs{o}$, $\un{yr}$ & $0.01$ & $0.01$& $0.01$\\
		First regrid time, yrs & 55 & 140 & 220\\
		Second regrid time, yrs & 900 & 1600 & 1800\\
		Third regrid time, yrs & 45 000 & 45 000 & 45 000\\
		\hline
	\end{tabular}
    \label{table_model_param}
\end{table}

While we choose the key explosion and CSM parameters in a similar way as \citet{2012APh....35..300T,2013A&A...552A.102T}, our setup differs in some aspects, in particular, we treat MF in our equations directly, leading to a full MHD numerical solution; the ISM structure in our simulations has a smooth transition between the CSM wind and the interstellar gas which allows us to see clearly the evolutionary transitions of SNRs. 

In practise, we run simulations for each of the three SN models from  Table~\ref{table_model_param} with the radial MF and then with the tangential MF. In this way, we have six models. However, it appears that hydrodynamic properties of the remnants (i.e. related to $\rho$, $p$, $v$) are the same either for the radial MF or for the tangential MF. Therefore, we analyse first the HD properties independently of the MF configuration and then look at the properties of the radial or tangential MF.

\begin{figure}
  \centering 
  \includegraphics[width=\columnwidth]{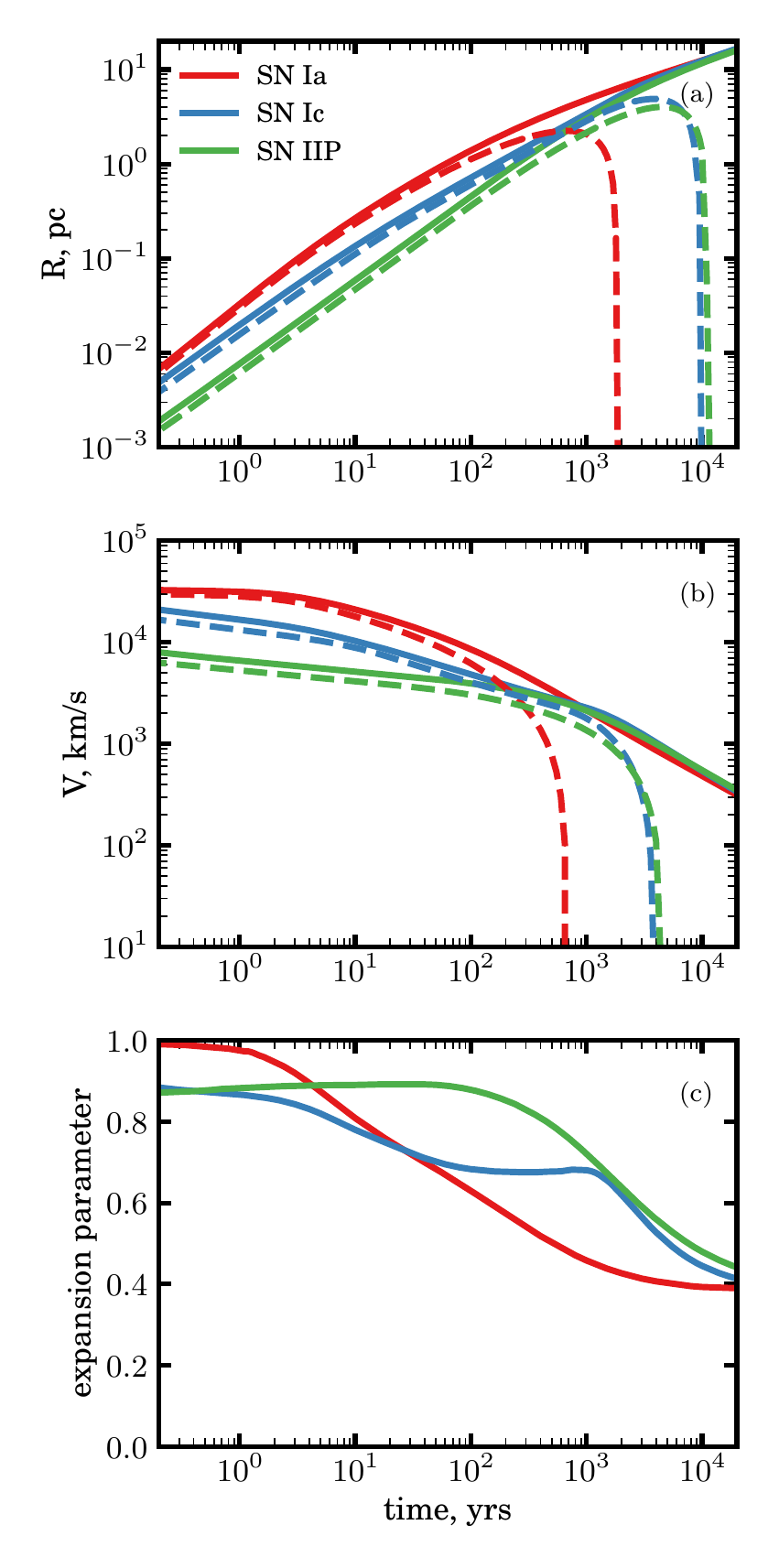}
  \caption{Dynamics of the remnants of different supernovae models 
  (represented by different colors): 
    evolution of the radius ({\bf a}), velocity ({\bf b}) of the forward (solid lines) and reverse (dashed lines) shocks as well as evolution of the expansion parameter $m$ for the forward shock ({\bf c}). 
  }
  \label{fig_SNI_all_shock_wave}
\end{figure}

\section{Overall dynamics}
\label{youngsnr:shock}

\subsection{Shock kinetics}
At first, we would consider properties of the flow motion. 
One of the informative characteristics is the expansion (or deceleration) parameter which is defined as 
\begin{equation}
m=\frac{d \log R}{d\log t} = \frac{t V}{R}.   
\end{equation}
It is the instantaneous power-law index in the representation  
$R\propto t^m$. Alternatively, it can be defined as the ratio of the true SNR age $t$ to the instantaneous expansion age $R/V$. 
The parameter $m$ is a useful indicator of the SNR evolutionary phase.
The solution of \citet{1982ApJ...258..790C,1985Ap&SS.112..225N} yields for $n>5$
\begin{equation}
    m=\frac{n-3}{n-s}.
    \label{youngsnr:mChevalier}
\end{equation}
In the Sedov solution for the strong adiabatic shock created by a point explosion 
\begin{equation}
    m=\frac{2}{5-s};
    \label{youngsnr:mSedov}
\end{equation}
for instance, $m=0.4$ for uniform ISM \citep{1959sdmm.book.....S}.

\begin{figure}
  \centering 
  \includegraphics[width=\columnwidth, trim={12pt 12pt 12pt 10pt},clip]{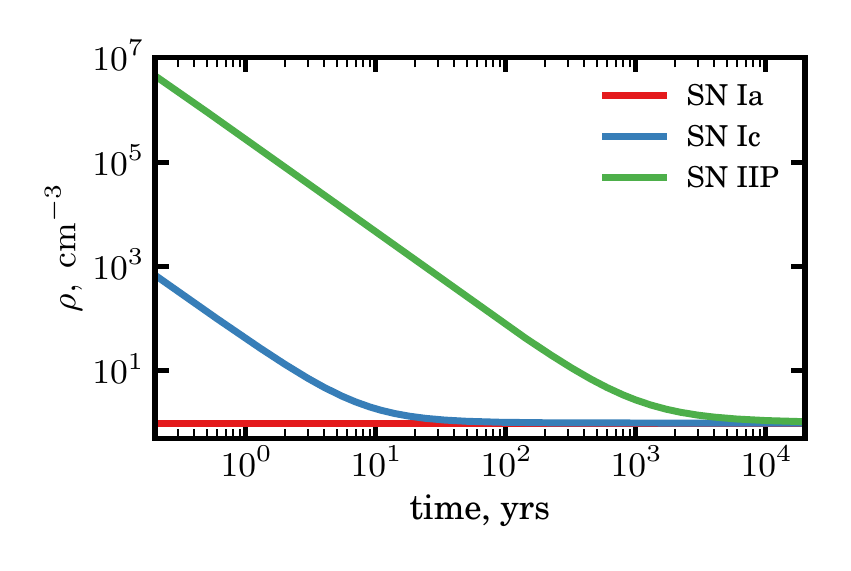}
  \caption{
    Pre-shock density in the ambient medium (i.e. at the forward shock location $R(t)$) as a function of SNR age $t$.
  }
  \label{fig_SNI_all_preshock}
\end{figure}

The red lines at Fig.~\ref{fig_SNI_all_shock_wave} show the evolution of the radius {$R$}, velocity {V} and expansion parameter {$m$} for the forward shock and the reverse shock for SNIa model which evolves in the uniform ambient medium. Well known features are visible. Namely, FS expands almost as $R\propto t$ (with $m\approx 1$) at the very beginning {(Fig.~\ref{fig_SNI_all_shock_wave}a)}. It decelerates with time, quite slowly till about $\sim 3\un{yr}$ and then more rapidly, approaching the Sedov regime with {$R\propto t^{2/5}$} ($m=0.4$) near the time when RS comes back and reaches the explosion point. 

Similar features are visible in the evolution of the remnants of CC SNe (Fig.~\ref{fig_SNI_all_shock_wave}, blue and green lines). Also in these cases the {FS radius starts to expand similarly to the Sedov solution} ($m$ approaches the value $0.4$) after RS returns to the explosion location (Animation \ref{animA1}: $9000\un{yr}$ for SNIc and $1000\un{yr}$ for SNIIP). 
These SNRs expand in the medium with the decreasing density $\rho\rs{o}\propto r^{-2}$, therefore, the transition happens somehow later (because FS decelerates slowly) compared to the SNIa model which evolve in the uniform medium {(Fig.~\ref{fig_SNI_all_shock_wave}a)}. 

The FS and RS velocities in our SNIc and SNIIP models follow \citet{1982ApJ...258..790C,1985Ap&SS.112..225N} solutions
\begin{equation}
 V\propto t^{(s-3)/(n-s)},
\end{equation}
that is $V\propto t^{-1/9}$ for $n=11$, $s=2$ (exactly as it is in Fig.~\ref{fig_SNI_all_shock_wave}b for the green line up to the age about $200\un{yrs}$) and $V\propto t^{-1/3}$ for $n=9$, $s=0$ (as the blue line between about few tens and $10^3\un{yrs}$).

\begin{figure}
  \centering 
  \includegraphics[width=\columnwidth, trim={4pt 10pt 7pt 10pt},clip]{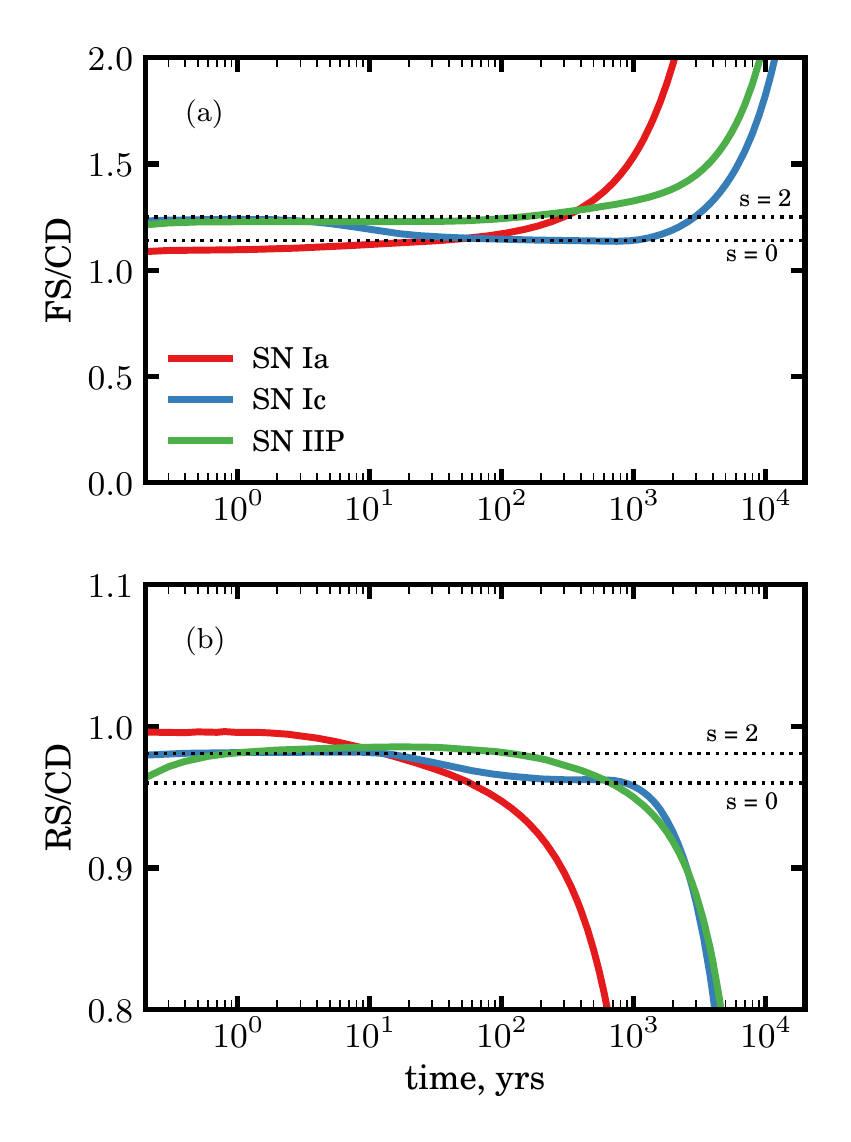}
  \caption{Evolution of the ratio of the radii derived from the simulations (solid lines): FS/CD ({\bf a}) and RS/CD ({\bf b}). 
     Dotted lines are theoretical predictions from \citet{1982ApJ...258..790C} model for $n = 9$ ($s = 2$ and $s = 0$). 
     Theoretical line for $n = 11$, $s = 2$ is not shown because of its proximity to the line for $n = 9$.
  }
  \label{fig_SN_all_r_ratio_shock_wave}
\end{figure}

\begin{figure}
  \centering 
  \includegraphics[width=\columnwidth, trim={4pt 13pt 6pt 6pt},clip]{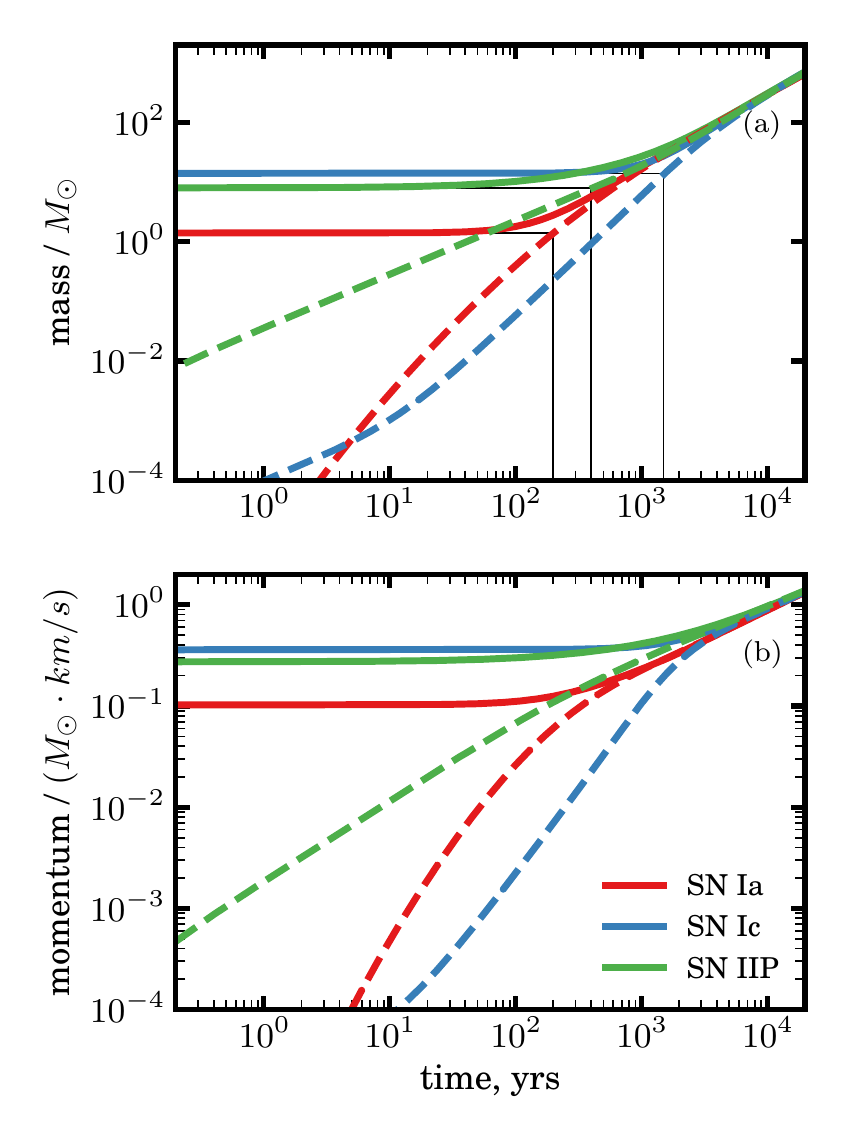}
  \caption{Evolution of the mass ({\bf a}) and momentum ({\bf b}). 
  Solid lines correspond to the whole SNR volume (integration from the explosion center to $R$). Dashed lines represent the mass and momentum of the shocked ISM gas only (between CD and FS). 
  {Thin black lines mark times when the swept-up mass becomes equal to the ejecta mass.}
  }
  \label{fig_SN_all_mass_momentum}
\end{figure}

The evolution of SNIa (shown by the red line  in Fig.~\ref{fig_SNI_all_shock_wave}c) during most of the early age consists in transition from the free expansion (the leftmost part with $m=1$) to the \citet{1959sdmm.book.....S} stage (the rightmost part with $m=0.4$). The CC SN models evolve first in the Chevalier ejecta-driven regime; the transition to the Sedov solution is shorter for them than in SNIa model but also takes considerable part of their ages.

\begin{figure*}
  \centering 
  \includegraphics[width=14.7truecm]{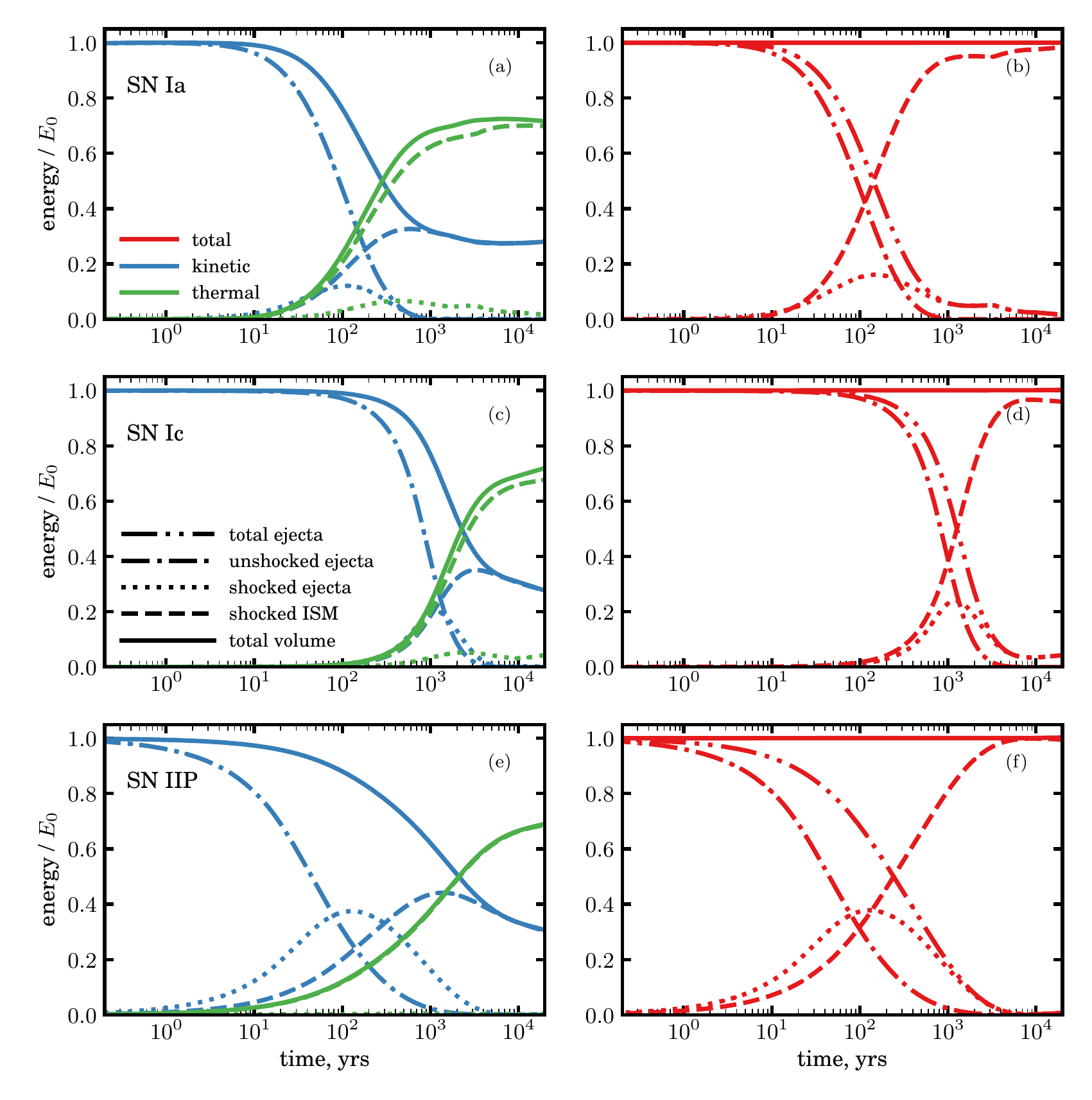}
  \caption{Evolution of the energy components in the remnants of different supernovae: Ia (top panels), Ic (middle panels), IIP (bottom panels). 
  The green dashed line on the plot (e) almost co-insides with the green solid line and the green dotted line is almost zero there.
  Plots on the left display the kinetic (blue lines) and thermal (green lines) energy. Plots on the right show the total energy (kinetic plus thermal). Solid lines correspond to the whole SNR volume, 
    dashed lines correspond to energies in the shocked ISM only (between CD and FS), dotted lines to the shocked ejecta (between RS and CD), the dash-dot lines to the unshocked ejecta (between zero to RS) and dash-dot-dot lines to the total energy in the ejecta (between zero and CD).}
  \label{fig_SN_all_energy_components}
\end{figure*}

Comparing Fig.~\ref{fig_SNI_all_shock_wave} with Fig.~\ref{fig_SNI_all_preshock}, one can see the impact of the ambient medium density on the shock dynamics. SNIa in our model evolves in a uniform medium, as shown by the red line in Fig.~\ref{fig_SNI_all_preshock}. {Instead, the two models of CC SNe assume power-law profiles for the ambient density (green and blue lines in Fig.~\ref{fig_SNI_all_preshock}), as the solutions of \citet{1982ApJ...258..790C,1982ApJ...259..302C}. The following features are evident.} 

\begin{itemize}
\item 
The green line (SNIIP model) in Fig.~\ref{fig_SNI_all_shock_wave}c is in a good agreement with the \citet{1982ApJ...258..790C} solution up to $\sim 200\un{yr}$ (equation (\ref{youngsnr:mChevalier}) gives $m=0.89$ for $n=11$ and $s=2$) and then tends to the \citet{1959sdmm.book.....S} one ($m=0.4$). 
In this model, SNR evolves till $10^3\un{yrs}$ in the power-law density profile without changes in its slope (green line in Fig.~\ref{fig_SNI_all_preshock}). 

\item 
Instead, in the case shown by the blue lines (SNIc model), FS `sees' the same decreasing density profile up to $\sim 10\un{yrs}$ only and then moves in a uniform medium (blue line in Fig.~\ref{fig_SNI_all_preshock}). Therefore, the parameter $m$ (blue line in Fig.~\ref{fig_SNI_all_shock_wave}c) demonstrates the transition from one (for $\rho\rs{o}\propto r^{-s}$ with $s=2$) to another solution (with $s=0$) of \citet{1982ApJ...258..790C}. 

\item 
{The transition time to the $s=0$ solution is about $50\un{yr}$. Indeed,} 
the deceleration parameter changes in this model ($n=9$) in accordance to the equation (\ref{youngsnr:mChevalier}): from $m=0.86$ for $s=2$ in the wind to $m=0.67$ for $s=0$ in the uniform ISM. 
\item Fig.~\ref{fig_SN_all_r_ratio_shock_wave} clearly shows that this is really the transition between the two \citet{1982ApJ...258..790C} solutions. In fact, the ratio of radii between FS/CD and RS/CD coincide with the values derived by \citet{1982ApJ...258..790C}; they are shown by the dashed lines Fig.~\ref{fig_SN_all_r_ratio_shock_wave}. SNIc model deviates from the \citet{1982ApJ...258..790C} solution at around $10^3\un{yr}$. 

\item 
As we have already pointed out, the green line in Fig.~\ref{fig_SN_all_r_ratio_shock_wave}b shows lower RS/CD ratio below $t=0.5\un{yr}$ because the shock runs at the beginning of simulations from the less dense ejecta into the CSM wind of the higher density.
\end{itemize}

We have also shown the SNIa case in Fig.~\ref{fig_SN_all_r_ratio_shock_wave} (red line) in order to have an idea of how the model with the exponential ejecta profile differs from the model with the power-law profiles. There is also a period with constant ratios between RS, CD, FS in the SNIa model; it lasts up to $\sim 3\un{yr}$ (Fig.~\ref{fig_SN_all_r_ratio_shock_wave}). 

\begin{figure}
  \centering 
  \includegraphics[width=\columnwidth, trim={4pt 13pt 6pt 6pt},clip]{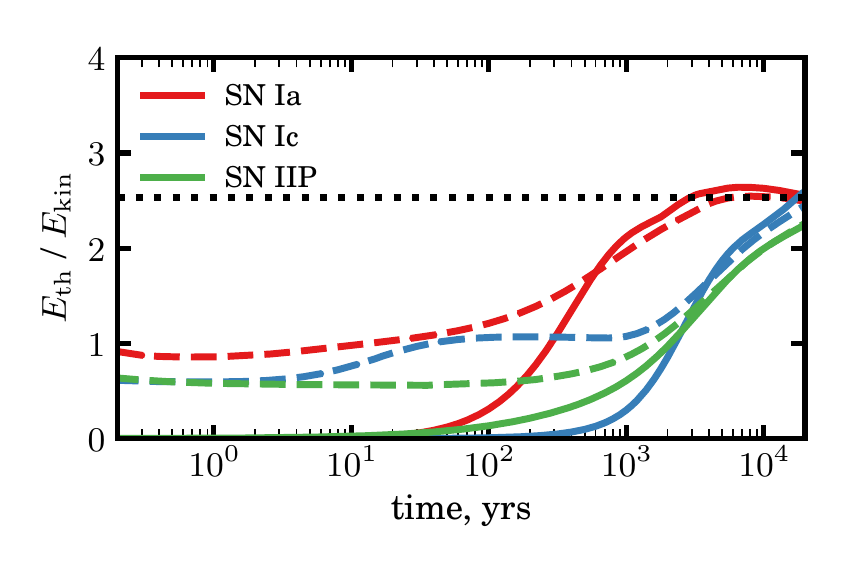}
  \caption{Evolution of the ratio of thermal and kinetic energy components. 
  Solid lines correspond to the whole SNR volume (integration from the explosion center to $R$). Dashed lines represent the energy components of the shocked ISM gas only (between CD and FS).
  Dotted black line corresponds to the theoretical value of $E\rs{th}/E\rs{kin}$ for the Sedov shock in the uniform medium.
  }
  \label{fig_SN_all_energy_ratio}
\end{figure}

Fig.~\ref{fig_SN_all_mass_momentum} shows components of SNRs' mass and momentum. The following properties may be noticed.
\begin{itemize}
\item 
The swept-up mass is almost unimportant up to the times when our CC SN models and the \citet{1982ApJ...258..790C} solution start to diverge ($\approx150\un{yr}$ for SNIIP and $\approx1000\un{yr}$ for SNIc, as from Figs.~\ref{fig_SNI_all_shock_wave} and \ref{fig_SN_all_r_ratio_shock_wave}). At these times, the ratio of the shocked ambient gas to the total mass within FS is $\approx 0.3$ and the swept-up mass (dashed lines) starts to modify the total SNR mass (solid lines). 

\item
In SNIa model (red lines), 
the swept-up mass rises to such a dynamically important level at around $100\un{yrs}$. At around this time, a rapid separation between the forward and reverse shocks initiates (Figs.~\ref{fig_SNI_all_shock_wave}a and \ref{fig_SN_all_r_ratio_shock_wave}). 
For this model, the parameter $m$ is already far in the transition from unity to the Sedov value $0.4$ (Figs.~\ref{fig_SNI_all_shock_wave}c). 

\item 
The swept-up mass equates to the mass of the progenitor star at $200\un{yr}$, $400\un{yr}$, $1400\un{yr}$ for SNIa, SNIIP and SNIc models respectively. In CC SN models, these times may approximately mark the end of the period when the \citet{1982ApJ...258..790C} solution may be approximately relevant.


\item
The evolution of momenta is quite similar to the evolution of masses (Fig.~\ref{fig_SN_all_mass_momentum}) {in our simulations} because the flow velocity is not very different between the models (Fig.~\ref{fig_SNI_all_shock_wave}b). 
\end{itemize}

It is important to note that the three solutions, despite of the initial differences in the ejecta structure, converge to the same evolutionary path after $\sim 7000\un{yrs}$ of evolution and become almost indistinguishable due to the fact that the dynamics is dominated by the interaction with the ambient environment. Of course one of the reasons is the usage of the same ISM density and explosion energy in the three cases. Actually, this fact proves that, at later times, the evolution of the remnant depends mainly on the properties of ISM through which the remnant propagates and on the energy of the explosion. {This result indicates the importance of considering a realistic ISM structure, in 3-D, for the simulations of SNRs of such or higher ages.}

To conclude, there is a period {of significant duration} when the SNR transits from the earliest phase toward the Sedov regime. At the end of this period, the interaction with ISM dominates over the ejecta structure (and the nature of the progenitor) in guiding the remnant evolution; SNR looses memory of the parent SN and the nature of the progenitor star.
{\citet{2019ApJ...877..136F} estimated that the remnant of an SNIa event loses memory of the progenitor after a couple of hundreds years of age. In the case of CC SNe the remnant is able to keep memory of the progenitor star through the ejecta structure for a few thousands of years \citep{2021A&A...645A..66O}. This result from numerical simulations is consistent with the observational evidence that the remnants of CC SNe are systematically much more asymmetric than those of SNIa (e.g.,~\citealt{2011ApJ...732..114L}).}

\subsection{Energy components}

Fig.~\ref{fig_SN_all_energy_components} gives a deeper insight into the transition from the early phase to the Sedov stage; it shows the evolution of the individual energy components. One can clearly see that the main distinction of the transition phase is redistribution of the explosion energy (which was initially in the form of the ejecta kinetic energy) into different components.
 
Explosion energy (blue dash-dot line) goes into the energy of shocked ejecta (dotted lines) and to the energy of the swept-up ISM material (dashed lines). The shocked ejecta energy is typically kinetic (left plots in Fig.~\ref{fig_SN_all_energy_components}). It increases initially to the maximum (at about the time when the kinetic energies of the unshocked ejecta and swept-up ISM equate at the level $\approx 0.25 E\rs{o}$) and then drops to the negligible value (that happens at around time when RS starts to move backwards in the ISM reference frame, see behavior of the density profiles at Animation \ref{animA1} and Figs.~\ref{fig_SNIa_full}-\ref{fig_SNIIP_full}: $600\un{yr}$ for SNIa, $4000\un{yr}$ for SNIc, $5000\un{yr}$ for SNIIP). 

With time, most of initial energy is tunnelled into the swept-up gas. 
The shocked ISM energy is partitioned between the thermal and kinetic components  (Fig.~\ref{fig_SN_all_energy_ratio}). Later on, the ratio between the components tends to the fixed value which corresponds to the \citet{1959sdmm.book.....S} solution, in particular, $2.54$ for the uniform ISM. The thermal to kinetic energy ratio in the swept-up material is also constant in the \citet{1982ApJ...258..790C} solution (green and blue dashed lines in Fig.~\ref{fig_SN_all_energy_ratio}): $E\rs{th}/E\rs{kin}\approx 0.6$ for $s=2$ and about the unity for $s=0$. 
The red line (SNIa model) demonstrates growth from $0.9$ to $1.1$ during the first $100\un{yr}$ and then increases to the Sedov ratio.

\subsection{Energy-conversion phase}
\label{youngsnr:transstage}

An important conclusion should be drawn from the above results. Namely, the transition time from the initial free-expansion phase (which -- for parameters in our models -- ends after $\sim 3\un{yr}$ for SNIa, at $\sim 300\un{yrs}$ for SNIIP and $\sim 10^3\un{yrs}$ for SNIc models) to the Sedov regime (when $m\approx 0.4$) is comparable or even larger than the free-expansion stage duration. Thus, there is a need to consider a new stage in a sequence of phases in the SNR evolution, with its own features. An approximate expression for FS motion during this stage has been developed by \citet{2017MNRAS.465.3793T} \citep[see also][]{1999ApJS..120..299T}. The idea of authors consists in matching the \citet{1982ApJ...258..790C} and \citet{1959sdmm.book.....S} self-similar solutions by a mathematical expression. In the present paper, we focus on the internal structures and physical properties of SNR in this intermediate stage. 

Main features of the transition stage are: i) energy conversion from the kinetic to the thermal form, ii) transfer of the explosion energy from the ejecta to the material of the shocked ambient medium.
All these are due to SNR expansion and the action of the reverse and forward shocks on the stellar ejecta and the swept-up material. 

Therefore, we propose to call this transition stage the `energy-conversion phase'. 

{We may formulate rather generally that the} transition ends when i) the swept-up material is dominant in the overall SNR dynamics by its mass, ii) the energy is almost completely deposited in the swept-up material, iii) the overall energy is partitioned between the thermal and kinetic components with the ratio {close} to the \citet{1959sdmm.book.....S} solution. This corresponds to ages: $\sim 10^3\un{yrs}$ for SNIa model, $\sim10^{4}\un{yrs}$ for SNIc and SNIIP models (Figs.~\ref{fig_SN_all_energy_components}, \ref{fig_SN_all_energy_ratio}). These features together with conditions of the negligible radiative losses and constant $E\rs{th}/E\rs{kin}$ ratio are the distinctive features of the following Sedov phase. In particular, the fixed value of the expansion parameter $m$ (0.4 for uniform medium) during the Sedov phase is a consequence of the constant ratio between the energy components in the swept-up material. In fact, the Sedov expression for the shock radius $R\propto (E\rs{o}/\rho\rs{0})^{1/5}t^{2/5}$ follows from the relation $M\rs{o}V^2/2=kE\rs{o}$ where $M\rs{o}(R)$ is the mass of ISM within $R$ and $k$ a constant.

\begin{figure*}
  \centering 
  \includegraphics[width=17.8truecm]{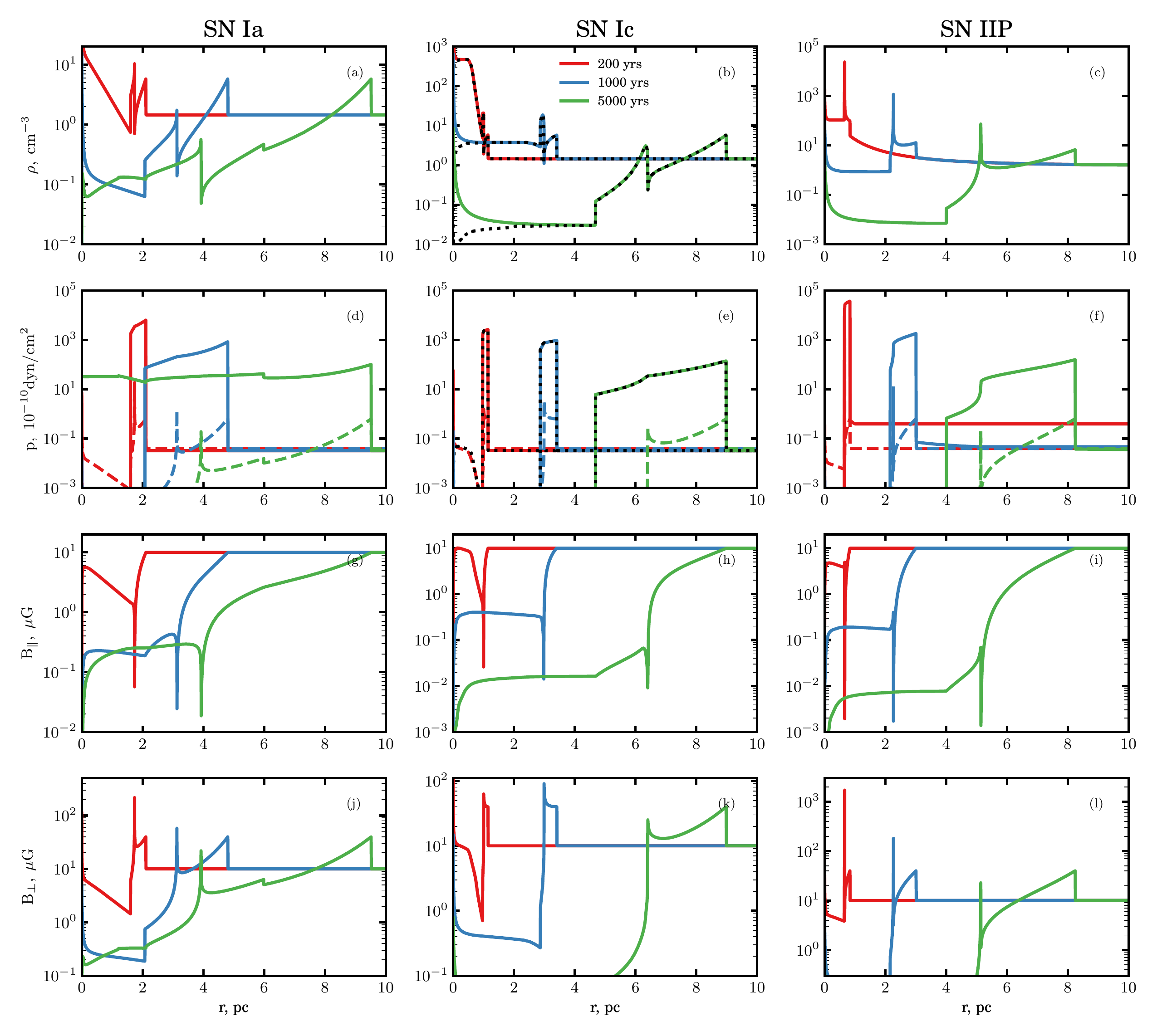}
  \caption{Radial profiles of density (top row), pressure (second row), radial MF (third row) and tangential MF (bottom row) for different SN explosion models at 200 yrs (red line), 1000 yrs (blue line) and 5000 yrs (green line). The dashed line on pressure plots corresponds to the tangential MF pressure. {All the lines for density and pressure are taken from the models with the tangential MF. Dots for the SNIc model represent the same profiles for the radial MF.}}
  \label{fig_SN_all_profiles}
\end{figure*}

\begin{figure}
  \centering 
  \includegraphics[width=8truecm]{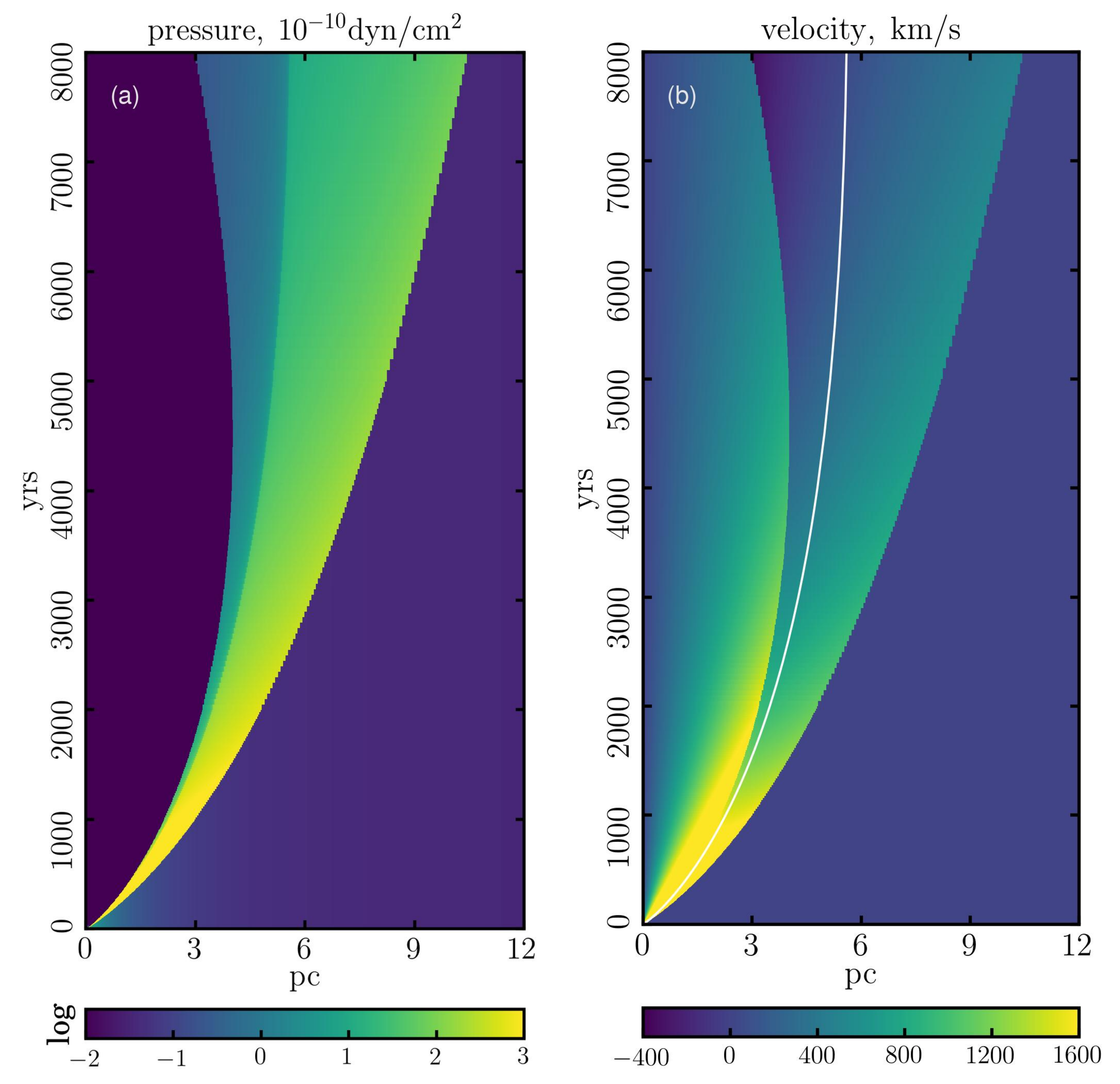}
  \caption{Evolution of the radial profiles of pressure ({\bf a}) and velocity ({\bf b}) for the SN IIP model. 
  The white line on the velocity plot corresponds to the position of the contact discontinuity. Units on the color scales are the same as in Fig.~\ref{fig_SN_all_profiles}}
  \label{fig_SNIIP_pv}
\end{figure}

\section{Evolution of MHD structure}
\label{youngsnr:interior}

\subsection{Hydrodynamics}

The profiles of hydrodynamic parameters $\rho$, $p$, $v$ in our simulations demonstrate the well known post SN explosion structure: two shocks, the reverse one moving in the ejecta and the forward one expanding into the ambient medium. The shocked ejecta and the shocked ambient material are in contact at CD. The unshocked ejecta stretches with the volume between the explosion center and RS. 
The evolutionary snapshots of the post-shock structure in our models are shown on  Fig.~\ref{fig_SN_all_profiles}.
These results are compatible with HD numerical models from \citet{1998ApJ...497..807D},  \citet{2012APh....35..300T,2013A&A...552A.102T} and others.

At the early times (red lines in Fig.~\ref{fig_SN_all_profiles}) the post-shock flow is characterized by the lower-density outer shell of heated ISM; the shocked ejecta forms {a 
shell behind} CD. With time, the density of the shocked ejecta becomes typically lower than the density of the swept-up gas and less important dynamically. It is interesting to note, that the density has a sharp downfall at CD from ejecta to ISM and then increases again from CD to FS in SNIa and SNIc models but {the density behavior differs (it has a local maximum at the CD)} in SNIIP case (Fig.~\ref{fig_SN_all_profiles} first row). The pressure is high only in the region which is perturbed by RS and FS (Fig.~\ref{fig_SN_all_profiles} second row).

\begin{figure}
  \centering 
  \includegraphics[width=7.3truecm]{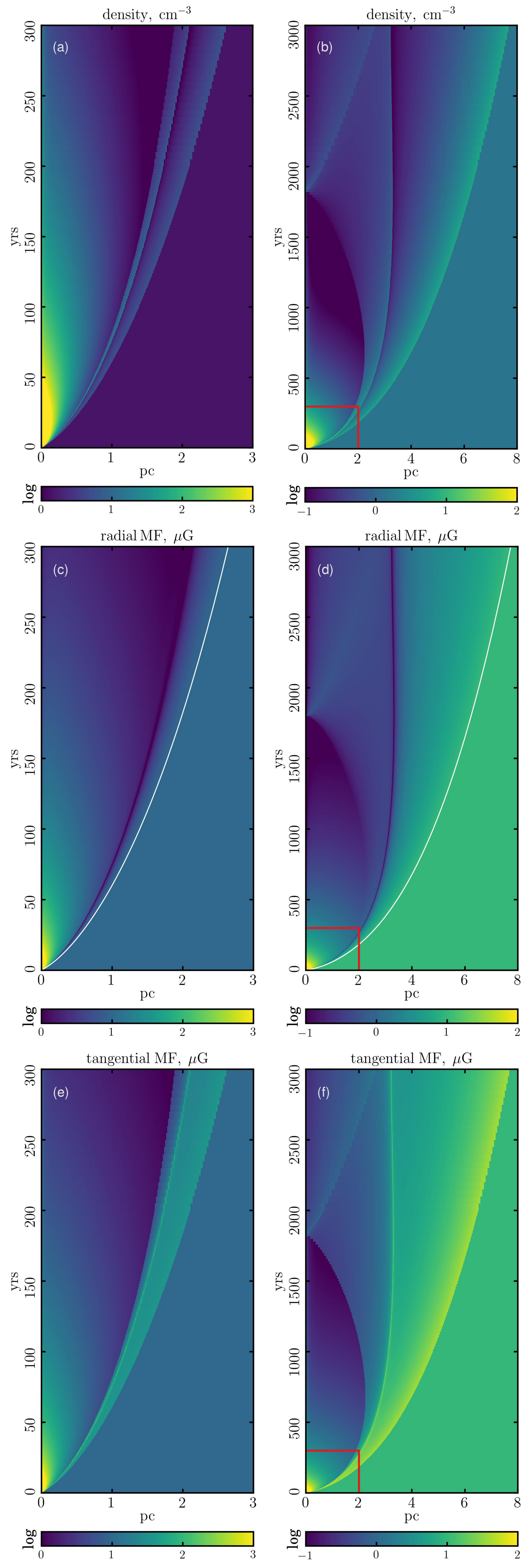}
  \caption{Evolution of the radial profiles for SNIa model: density ({\bf a-b}), radial MF ({\bf c-d}) and tangential MF ({\bf e-f}). Red boxes on the right panels correspond to the whole scans displayed on the left panels.
  {White lines in the panels c and d mark the FS position.}}
  \label{fig_SNIa_full}
\end{figure}

The space-temporal scans of post-shock flows are useful to catch the evolutionary properties. As an example, Fig.~\ref{fig_SNIIP_pv} demonstrates the evolution (vertical axis) of the pressure (Fig.~\ref{fig_SNIIP_pv}a) and gas velocity (Fig.~\ref{fig_SNIIP_pv}b) profiles for SNIIP model. The pressure in the region between RS and FS decreases with expansion of the region. The CD separating the two shocked media are clearly seen on the pressure plot. Instead, CD is not quite prominent on the velocity plot because there is no sharp change in the fluid velocity at CD. Velocity decreases with $r$ from CD to FS at early times and increases later. It grows linearly (free expansion) from zero to RS all the time until RS reaches the point $r=0$. Such an event is clearly visible on the $r-t$ scans for density in SNIa model (Fig.~\ref{fig_SNIa_full}b): this happens at around $1800\un{yrs}$. 

Animations \ref{animA1}, \ref{animA2} give useful insight into the evolution of the MHD structure. Also here, one can see how the reverse shock in all the models 
reaches the origin, bounces back and runs through the whole SNR towards FS (corresponding feature is visible on green lines in Fig.~\ref{fig_SN_all_profiles} a,d around 6 pc). 
About that time ($1800\un{yrs}$ for SNIa, $9000\un{yrs}$ for SNIc, $11000\un{yrs}$ for SNIIP), the FS expansion parameter is already quite close to $0.4$ (Fig.~\ref{fig_SNI_all_shock_wave}c), and the post-shock distribution of $\rho$ and $p$ tend to Sedov profiles, first of all near the shock (Animation \ref{animA2} left column). 

\begin{figure}
  \centering 
  \includegraphics[width=7.63truecm]{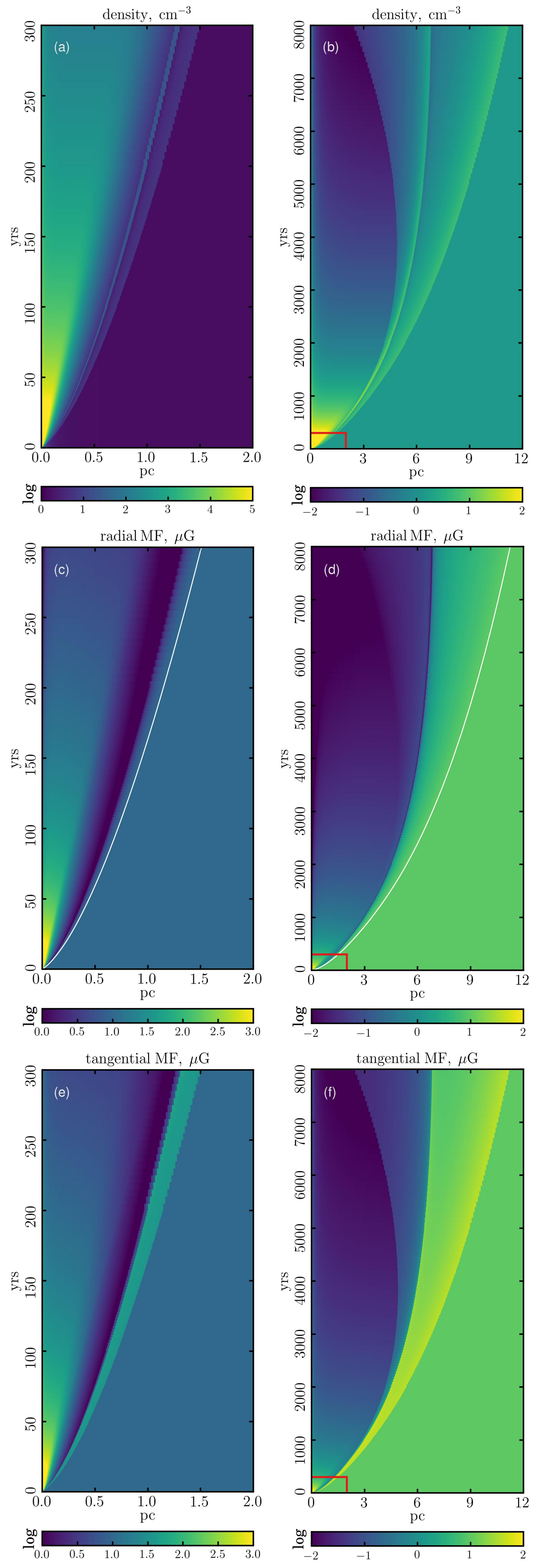}
  \caption{The same as in Fig.~\ref{fig_SNIa_full} for SNIc model.}
  \label{fig_SNIc_full}
\end{figure}
\begin{figure}
  \centering 
  \includegraphics[width=7.63truecm]{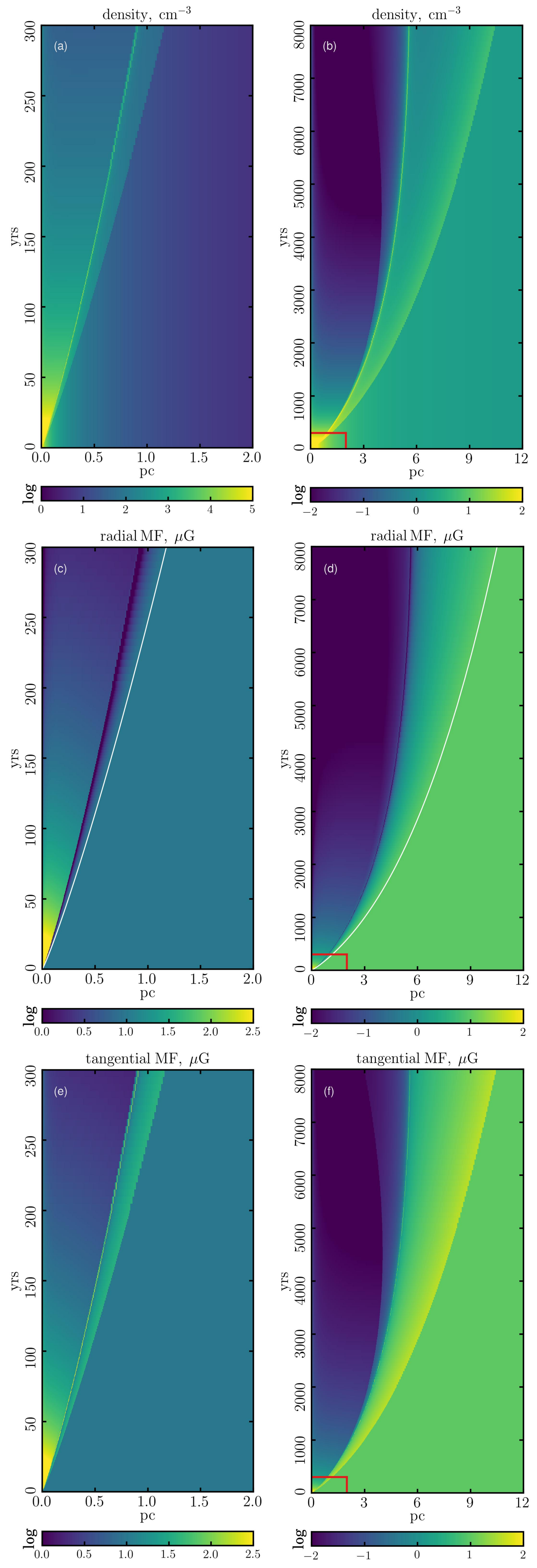}
  \caption{The same as in Fig.~\ref{fig_SNIa_full} for SNIIP model.}
  \label{fig_SNIIP_full}
\end{figure}
\begin{figure*}
  \centering 
  \includegraphics[width=14.7truecm]{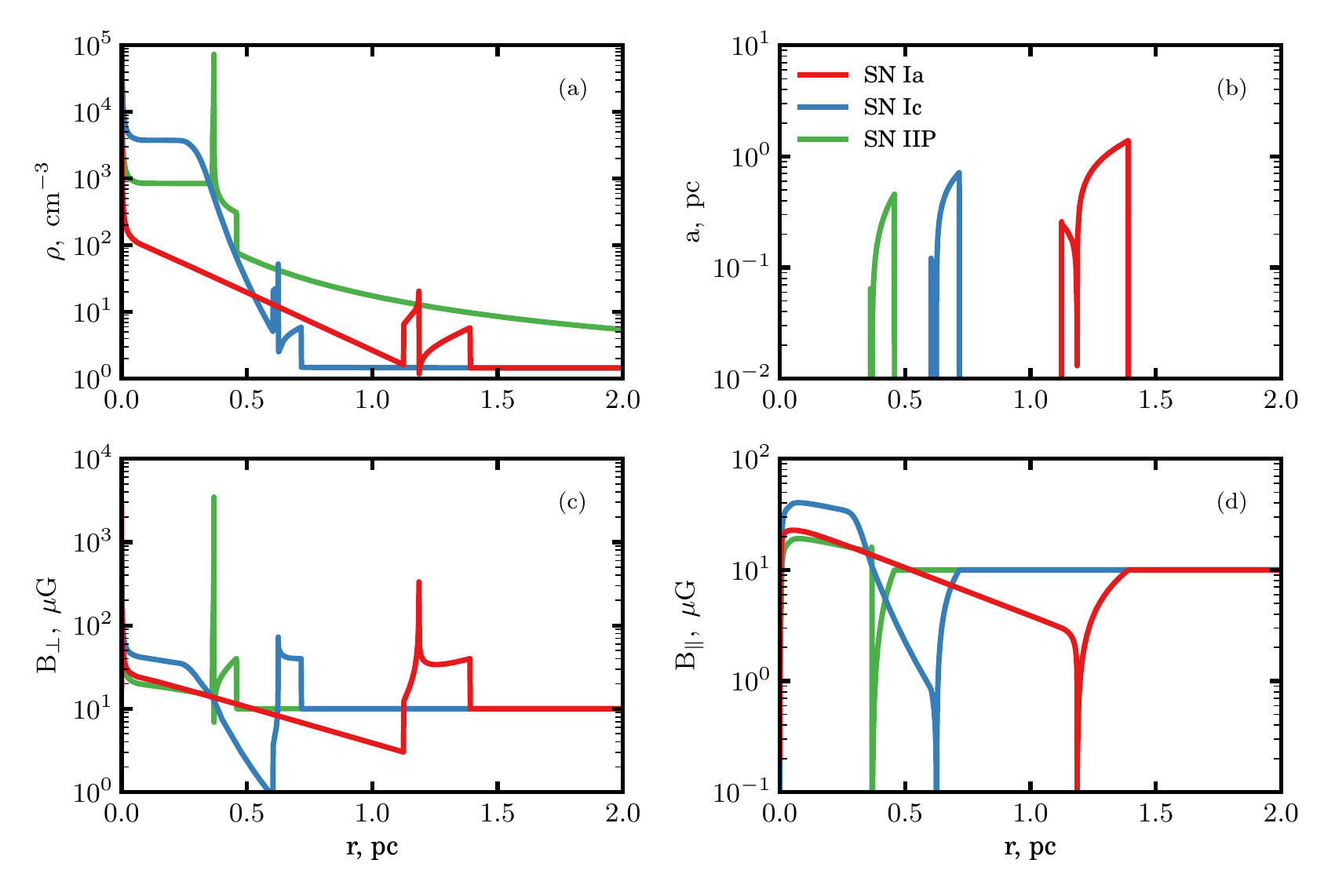}
  \caption{Radial profiles of density ({\bf a}), Lagrangian coordinate ({\bf b}), perpendicular ({\bf c}) and parallel ({\bf d}) MF strength at 100 yrs for three SNR models.}
  \label{fig_SN_all_MF_100}
\end{figure*}

An interesting effect is clear from the  $r-t$ scans, namely, CD is almost stationary after some time: it stays at about 3 pc for SNIa (Fig.~\ref{fig_SNIa_full}), 6 pc for SNIc (Fig.~\ref{fig_SNIc_full}), 5 pc for SNIIP model (Fig.~\ref{fig_SNIIP_full}). This happens at around 600, 3000, 4000 yr respectively, i.e. when kinetic energy in the (unshocked and shocked) ejecta approaches to zero (see Fig.~\ref{fig_SN_all_energy_components}), so the  ejecta is not essentially expanding any more. CD moves somehow at later times mostly due to a push from the reflected RS. After these times the flow structure is more complicated due to the multiple waves \citep{1988ApJ...334..252C} propagating back and forth; in addition, the density of ejecta is typically much lower than the density of the swept-up material; so, the flow tends to the Sedov-like profiles, Animation \ref{animA1}.

Animations \ref{animA2} demonstrates that the profiles in the CC SN models are self-similar, as in \citet{1982ApJ...258..790C} solution. Self-similarity lasts up to $\approx 100\un{yr}$ for SNIIP model (green line). As to the SNIc model (blue line), it has the same property; however, since the FS moves from the ambient medium with $s=2$ into the medium with $s=0$ around $10\un{yr}$ (Fig.~\ref{fig_SNI_all_preshock}), i.e. switches between two solutions, the property of the self-similarity is hardly to notice at the beginning. 
{SNIc} are not self-similar after about $10^3\un{yr}$. 
Deviation from the self-similarity around these times is visible also from Fig.~\ref{fig_SN_all_r_ratio_shock_wave}. 
{Also, Fig.~\ref{fig_SNI_all_shock_wave}c demonstrates} that the deceleration parameter $m$ diverges from the \citet{1982ApJ...258..790C} values around the same times; thus, again, the parameter $m$ is a good marker of the SNR evolutionary stage. 

\subsection{Magnetic field}

The most interesting for us is the behaviour of the magnetic field.
We have performed two independent runs for each of our three models, one with the radial and another with the tangential MF. 
It is worth noting, that, even during the earliest times in SNR evolution, the pressure of MF in the FS-shocked gas is a few orders of magnitude lower than the thermal pressure (Fig.~\ref{fig_SN_all_profiles}d-f). This is the reason why the magnetic energy is not shown in Fig.~\ref{fig_SN_all_energy_components}. 

At initialization of our simulations, MF in the ejecta was taken to contribute locally only 10\% to the total pressure. At some time the magnetic field in evolving SNR will be dynamically relevant somewhere in the ejecta, because the adiabatic cooling of the rapidly expanding material very quickly reduces the internal temperature. Such an effect is visible in our simulations as well, namely, on the spatial profiles of density and MF  (Fig.~\ref{fig_SN_all_profiles}): the distributions either rapidly rise ($\rho$ and $B_\perp$) or decline ($B_\|$) as $r\rightarrow 0$ for the small $r$. Plasma becomes magnetically dominant just around the center of SNR that is unimportant for the overall dynamics.

In other words, if magnetic energy in the supernova progenitor was low comparing to the thermal one then the dynamical role of magnetic fields in SNR is small enough and its influence on the HD structures may be neglected as long as SNR evolves adiabatically. Instead, the tangential component of MF is a critical component determining the flow evolution in the post-adiabatic era
\citep{2016MNRAS.456.2343P,2018MNRAS.479.4253P}.

Fig.~\ref{fig_SN_all_profiles} demonstrates also that the MF orientation in respect to the shock normal does not create differences in the hydro-dynamics for our models (cf. dots for the radial MF and lines for the tangential MF for the SNIc model). The differences are only around the center where MF pressure dominates the thermal one, as described above.

At early stages, the structure of the tangential MF (Fig.~\ref{fig_SN_all_profiles} last row) demonstrates the two distinctive regions (RS-CD and CD-FS). Firstly, it has a jump at the (perpendicular) shock. Secondly, it peaks at a contact discontinuity where the flow density experiences a jump. The MF strength in the shocked ejecta monotonically grows with the radius. In the shocked ISM, it has concave shape in SNIa and SNIc models (Fig.~\ref{fig_SN_all_profiles} j,k) while increases in SNIIP, unlike to the concave shape of the density distribution in this model (cf. plots l and c in Fig.~\ref{fig_SN_all_profiles}). With time, the tangential MF behaves in a quite similar fashion to the density (Figs.~\ref{fig_SNIa_full},\ref{fig_SNIc_full},\ref{fig_SNIIP_full}).

The radial MF strength $B\rs{\|}$ is below the ambient value (Fig.~\ref{fig_SN_all_profiles}g-i), in contrast to the tangential one $B\rs{\perp}$  whose strength is above the pre-shock value in most of the swept-up material during all the time. The rapid fall of the MF strength downstream of the parallel shock has the same reason as in the post-adiabatic flows \citep[Sect.3.3 in][]{2016MNRAS.456.2343P}. In terms of Lagrangian coordinate $a$, the continuity equation reads \citep{1999A&A...344..295H}
\begin{equation}
 \rho(a,t)=\rho(a,t\rs{i})\left(\frac{a}{r}\right)^2\left(\frac{dr}{da}\right)^{-1}
\end{equation}
where $a$ is defined as $a\equiv R(t\rs{i})$, $t\rs{i}$ is the time when a given fluid element is shocked. MF components in the fluid element evolve as
\begin{equation}
 B\rs{\perp}(a,t)=B\rs{\perp}(a,t\rs{i})\left(\frac{r}{a}\right)\frac{\rho(a,t)}{\rho(a,t\rs{i})},
 \label{MFtang}
\end{equation}
\begin{equation}
 B\rs{\|}(a,t)=B\rs{\|}(a,t\rs{i})\left(\frac{a}{r}\right)^2.
  \label{MFrad}
\end{equation}
It is clear from these equations i) that the density directly affects the shape of the tangential MF strength profile ($\rho(a,t)$ is present in the equation~\ref{MFtang} but is not in the equation~\ref{MFrad}), ii) that the SNR expansion lowers the radial MF component (the Lagrangian coordinate $a$ is always smaller than the Eulerian $r$ after the shock passage through the fluid element $a$; therefore, $B(t)\leq B(t\rs{i})$ in a given fluid element for $t\geq t\rs{i}$), iii) the dependence $r(a)$ determines not only the profiles of HD parameters \citep{1999A&A...344..295H} but also the spatial distribution of $B$. This property could be useful: if the relation $r(a)$ is known (or approximated) then one may restore the distribution of all the MHD parameters. Fig.~\ref{fig_SN_all_MF_100} {illustrates for the first time how the density, the Lagrangian coordinate, and the MF components are related one to another in young SNRs}. 

There are observational evidences that polarization maps in young SNRs rather prefer the radial MF orientations \citep{1976AuJPh..29..435D,2015A&ARv..23....3D}. However, our simulations, in agreement with equations (\ref{MFtang})-(\ref{MFrad}), show that the radial MF component may not dominate over the tangential one in 1-D models of young SNRs, if these two components were of the same strengths initially. In 3-D simulations the radial MF might prevail in `fingers' developed due to Richtmyer-Meshkov instability \citep{2012ApJ...749..156O}. Recent 3-D MHD simulations performed by \citet{2019A&A...622A..73O} shows that the radial MF may appears due to instabilities at CD from an ambient magnetic field which was tangential initially (Fig.~3 in this reference). So, the role played by HD instabilities is crucial to let the radial component of the magnetic field to emerge in young SNRs.

\section{Discussion}
\label{youngsnr:discuss}

\subsection{Phases in evolution of SNR}

The early stage in the evolution of SNR is dominated by the  free-expansion of the gas, which is characterized by the linear dependence of the unshocked ejecta velocity: $u(r)=r/t$. 
In case when the density in the initial ejecta and ambient medium follow the power law, there are analytical expressions for shock motion as well as numerical and simplified analytical solutions for the internal structure   \citep{1982ApJ...258..790C,1982ApJ...259..302C,1985Ap&SS.112..225N}. 
The term `free expansion' in respect to the evolutionary stage should not be mixed up with the free-expansion of a shock when the shock speed is just  $V=R/t$, with the deceleration parameter $m=1$. In CC SN models the parameter $m$ corresponds to the value given by the equation (\ref{youngsnr:mChevalier}) already from the very beginning.

Half a century ago, 
\citet{1970IAUS...39..229W,1972ARA&A..10..129W}
has suggested to divide the life-long evolution of SNR into the following 
stages: free-expansion -- Sedov -- radiative. At the end, when the shock speed slows down to the speed of ISM random motions, SNR merges into ISM.

It was shown that a new phase need to be introduced into this sequence between the Sedov and the (fully) radiative stages because the duration of the transitional period is comparable with the SNR age at the end of Sedov phase \citep{2005JPhSt...9..364P}. The term to mark this phase -- `post-adiabatic' -- was suggested by \citet{2007MPLA...22.2617T}. Hydro-dynamical properties of the post-adiabatic SNRs were studied already before by \citet{1988ApJ...334..252C} and \citet{1998ApJ...500..342B}. Magneto-hydrodynamics of SNR during this transition was considered by \citealt{2016MNRAS.456.2343P,2018MNRAS.479.4253P}. 

It is worth noting that SNR evolve all the time till the end of Sedov stage in the adiabatic regime.  Therefore, the term 'adiabatic' is appropriate to refer all the time from the supernova explosion to the prominent radiative losses, not only to the Sedov phase.

The need for a separate consideration of another transitional period,  between the earliest ejecta-dominated stage and the Sedov one, was demonstrated by \citet{1999ApJS..120..299T} and \citet{2017MNRAS.465.3793T}, by considering differences in the the shock motion in the two stages. In the present paper, we analyse MHD properties of SNR during this period and 
suggest to call the transition the energy-conversion phase.

Thus, the idealized sequence of SNR evolutionary stages is as follows: 
free expansion of gas (or ejecta-dominated) -- energy-conversion -- Sedov-Taylor -- post-adiabatic -- (fully) radiative. Each of this stages has distinctive properties that obliged us to update the initial scheme of SNR evolution.

\begin{figure}
  \centering 
  \includegraphics[width=8truecm]{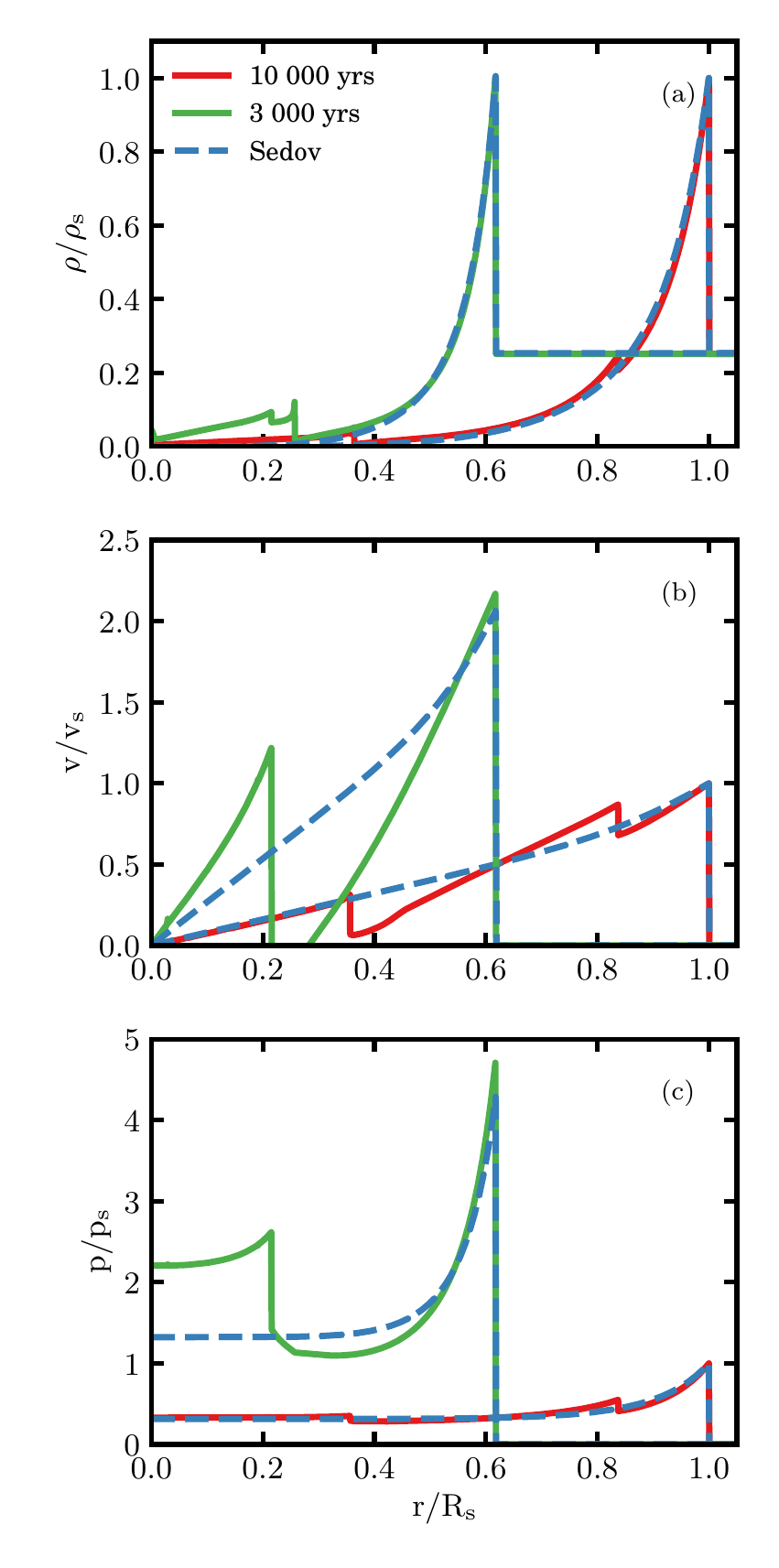}
  \caption{Radial profiles of density ({\bf a}), velocity ({\bf b}) and pressure  ({\bf c}) at $3\,000$ yrs (green lines) and $10\,000$ yrs (red lines) for our SNIa model compared to the Sedov profiles (dashed lines). The profiles are normalized to the post-shock values at $10\,000$ yrs}
  \label{fig_SNIa_profile_vs_sedov}
\end{figure}
\begin{figure}
  \centering 
  \includegraphics[width=\columnwidth, trim={12pt 12pt 12pt 10pt},clip]{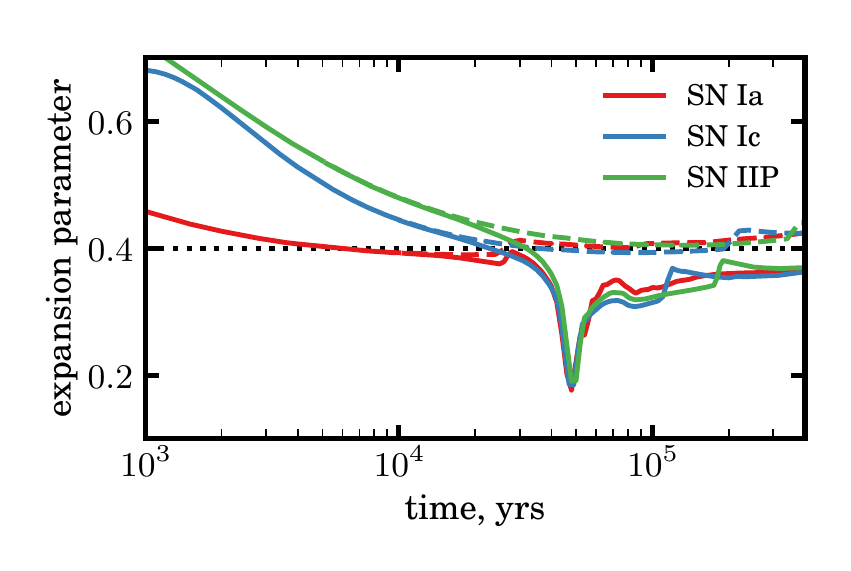}
  \caption{
    Evolution of the expansion parameter $m$ for the forward shock from the early stages up to a well-developed radiative stage.
    Dashed lines correspond to the same models but in completely adiabatic regime when the radiative losses are switched off artificially.
  }
  \label{fig_SNI_all_expansion_v2}
\end{figure}

\subsection{On-set and duration of the Sedov stage}

\citet{1959sdmm.book.....S} solution assumes a strong point explosion. The first condition (huge energy deposition) is obviously valid for SNe. The second means that the size and internal structure of the exploded material is negligible comparing to the perturbed region. Initial conditions in our simulations (and in real situations) are not point-like because the simulations start from the spatial distributions of parameters in the ejecta. It is well known that SNR therefore enters the Sedov regime later when SNR dynamics is dominated by the swept-up material.
It was already mentioned at the end of Sect.~\ref{youngsnr:transstage} that onset of the Sedov stage in our models happens at $\sim 2\E{3}\un{yr}$ (SNIa) and $\gtrsim10^4\un{yr}$ (SNIIP and SNIc) when the deceleration parameter $m$ is close the Sedov value (Fig.~\ref{fig_SNI_all_shock_wave}c).
\citet{1977PASJ...29..813I} suggested the criterion that the adiabatic shock approaches the Sedov phase when the swept-up material reaches {ten times the mass of the} SN ejecta. In our simulations {this point would be reached at $10^3$, $4\E{3}$ and $7\E{4}$ yrs, respectively (Fig.~\ref{fig_SN_all_mass_momentum}a), a bit earlier than the actual beginning of the Sedov phase. Hence we propose that a more reliable criterion could be the accumulation of $\approx 20-30\ M_\mathrm{ej}$.}

What about the profiles of the flow parameters? How close they are to the 
self-similar solution? The numerical profiles in SNIa model are compared to the \citet{1959sdmm.book.....S} solution in Fig.~\ref{fig_SNIa_profile_vs_sedov}. One can see that the profiles are not exactly the same even around $3000\un{yr}$; however, they are approaching analytical profiles, at first around FS, where the swept-up material is located and where most of the emission comes from.

The Sedov description is valid for the adiabatic shock. {Radiative cooling of the plasma is represented by the term $L=-n_en_H\Lambda(T)$ in the energy conservation equation, where $\Lambda$ is the cooling coefficient \citep[see Eq.~1 in][]{2018MNRAS.479.4253P}}. When is becomes relevant, the post-adiabatic stage begins, and a smooth decrease of the parameter $m$ commences (Fig.~\ref{fig_SNI_all_expansion_v2}, around $3\E4\un{yrs}$). This is a known property but what we would like to note is that the duration of the Sedov stage is quite short for our CC SN models (blue and green solid lines on this figure). In fact, it looks from this plot that the energy-conversion stage rather directly transforms into the post-adiabatic one. In order to check if this is peculiar property of the models considered, we have made other runs with different initial parameters and noticed that it is rather difficult to find a model with clear and well extended Sedov phase. We made another check by calculating the same models without radiative losses {(i.e. with $L=0$)}. In this case, all the models run in the Sedov regime for a long time (Fig.~\ref{fig_SNI_all_expansion_v2} dashed lines). 

The post-adiabatic decrease of the deceleration parameter $m$ happens in the same time for all our models. The reason is that the radiative losses depend on the post-shock temperature {$L\propto \Lambda(T)$, which in turn} is proportional to $V^2$. The shock speed $V$ in all our models are almost the same after $\sim 10^3\un{yr}$ (Fig.~\ref{fig_SNI_all_shock_wave}b). 

{The onset of radiative losses is tied to a certain FS speed. The rate of radiative losses peaks around $T\approx 2\E{5}\un{K}$  \citep[see Fig.~1 in][]{1998ApJ...500..342B}. The shock speed and the post-shock temperature are related as $V\approx 0.3 T^{1/2}\un{km/s}$. Therefore, the radiative losses are fastest when the shock speed is $V\approx 130\un{km/s}$. 
Since losses are related to the shock speed,} some deviations from the ideal scheme of SNR evolution may happen. In fact, the shock speed depends on the pre-shock density $V\propto \rho\rs{o}(R)^{-1/2}$ \citep{1999A&A...344..295H}. 
If the ISM density has local steepness $s(R)=-d\ln\rho\rs{o}/d\ln R>3$, i.e., density decreases fast enough, then FS accelerates. Therefore, if in some place the density begins to drop with $s>3$, the speed of the shock rises here and the post-adiabatic SNR may even return to the adiabatic regime again.

\subsection{{Are some effects due to 1-D geometry?}}
\label{youngsnr:discuss1D}

{In the present study, we decided to perform 1-D simulations instead of 3-D ones because they may be done with a resolution that is high enough to catch physical effects on small scales, to accurately resolve the shock locations, and to study the evolution and motion of disturbances. 3-D calculations would be desirable {but, at present, they may be performed on a grid not larger than (a few 1000)$^3$ points. 3-D simulations with a resolution which can be reached in 1-D runs (hundreds of thousands grid points) are prohibitive.} 

1-D simulations have weaknesses as well. In particular, they cannot treat in a self-consistent way multi-dimensional effects. If an SNR is asymmetric due to either an aspherical explosion, instabilities at tangential interfaces, or 3-D non-uniformity of the ISM, {then only 3-D simulations may model the object.} 
In some cases (if SNR is not highly asymmetric), one can split this object in a number of sectors and consider the overall SNR evolution in the sector approximation under the assumption that the sectors evolve independently. The technique still ignores HD or MHD instabilities that essentially require three dimensions to develop, in particular during their non-linear phase. Such instabilities introduce perturbations in the radial distributions of parameters and may relocate the position of CD and RS (clear examples are the `fingers' from the Rayleigh-Taylor instability which 1-D calculations can not describe).
{Furthermore, in case of a strongly inhomogeneous medium, the effect of interaction of the blast wave with the clouds (for instance) affects the remnant interior even in other sectors \citep{2020arXiv201208017U}.} 

In subsequent work, we intend to compare the present 1-D results to 3-D MHD simulations. We emphasize here that 1-D simulations -- which allow one to deeply resolve a system -- are an essential step toward the ``real'' situation because they are able to accurately catch the major properties of magnetized fluids with shocks.

In the following we discuss which effects in our simulations could possibly arise on account of the reduced dimensionality. }

\subsubsection{{Oscillations of the parameter $m$}}

{One debatable feature is seen in the behavior of the shocks after $\approx 50\,000$ yrs on Fig.~\ref{fig_SNI_all_expansion_v2} (solid lines). The sharp increase from the minimum and the repeated rising and falling of the deceleration parameter, $m$, is well known in 1-D HD simulations of radiative shocks \citep{1988ApJ...334..252C, 1998ApJ...500..342B}. In particular, \citet{1998ApJ...500..342B} analysed different setups and demonstrated that this is a physical effect and not the artifact of simulation features. Our 1-D MHD simulations of post-adiabatic shocks \citep{2016MNRAS.456.2343P,2018MNRAS.479.4253P} confirm this behaviour in magnetized systems, although the oscillations may be suppressed by a sufficiently strong MF \citep[see Fig.~4 in][]{2016MNRAS.456.2343P}. The physics behind the oscillation of $m$ is explained by \citet[][p.~2349]{2016MNRAS.456.2343P}. 
We believe that this effect results from 1-D geometry of simulations, because the flow is `locked' in one dimension without the possibility to move non-radially. The behavior of $m$ in a 3-D system is expected to be more smooth, without oscillation after the minimum, in agreement with the analytical solution \citep[][see Fig.~1 there]{2004A&A...419..419B}.}

\subsubsection{{RS bounce}}

{In our 1-D simulation we may see, especially from  Animation \ref{animA1}, how the RS reaches the explosion origin (at $t=1800\un{yr}$ for the SNIa model), bounces back,  moves out, and interacts with the CD. In the case of our SNIa model the interaction between the reflected RS and the CD happens around 3400 yrs (see the green lines on Fig.~\ref{fig_SN_all_profiles}). That interaction leads to a series of smaller waves running in different directions along the radius. In reality there will be some asymmetry that distributes that over time and leads to many shocklets moving non-radially.} {Could the sharp RS bounce at the center disappear if the three dimensions are properly considered?}

{In order to see whether the RS bounce is an artifact of the reduced dimensionality, we look for the effect in 3-D simulations. To this end, we considered a 3-D MHD simulation which describes the blast wave originating from a SN explosion that propagates into a wind-blown bubble surrounded by a thin dense, cool shell that formed from the interaction between the wind of the
progenitor star and the surrounding medium (\citealt{2009MSAIS..13...97O}). The evolution of the blast wave and ejecta were modeled by numerically solving the time-dependent MHD equations of mass, momentum, energy, and magnetic flux conservation in a 3D Cartesian coordinate system $(x,y,z)$. The model takes into account the effects of an organized ambient magnetic field, of the magnetic-field-oriented thermal conduction (including the effects of heat flux saturation; \citealt{2008ApJ...678..274O}) and of the radiative losses from an optically thin plasma. The initial condition represents the remnant of a $15 M_{\odot}$ star that exploded about $10^5$ years after the onset of the red supergiant (RSG) phase. The simulation assumes an explosion energy of $10^{51}$ ergs (mostly as kinetic energy). The ambient medium is modeled as a RSG wind-blown bubble of about 13 pc radius, surrounded by a cavity wall, namely a thin dense shell formed due to the accumulation of wind swept-up material at the interface with the ISM (\citealt{2005ApJ...630..892D}). For the sake of simplicity, the initial magnetic field is assumed to be uniform in the whole domain with strength $B = 1 \mu$G and parallel to the $x$ direction; this conditions lead to a ratio of hydrodynamic to magnetic pressure, $\beta$, much larger than 1 everywhere in the spatial domain.} 

{The simulation shows that, soon after the SN explosion, the blast wave expands in the bubble through the wind of the progenitor star with very high velocity. The blast wave reaches the cavity wall after about 2500 yrs. After this time, the forward shock propagates with much lower velocity through the high density medium of the cavity wall and, as a consequence, becomes radiative. The interaction with the dense material of the cavity wall drives a reflected shock, which propagates inward through the shocked wind and the ejecta. The reflected shock reaches the center of the remnant (which coincides with the center of the explosion) after $\approx 18000$ yrs. After the refocusing the shock bounces back and propagates outward reaching the cavity wall where it is reflected again backward.}

{The complete evolution of the spatial distribution of mass density is shown in the Animation \ref{animA3}. The movie shows that the shock front propagates back and forth between the cavity wall and the center of explosion several times. We note that the almost perfect focusing of the reflected shock at the center of the remnant and the shock bouncing are the result of the idealized geometry adopted for the pre-SN CSM. The blast wave reaches the cavity wall at all angles at the same time because the dense shell of the cavity wall is modeled as a sphere centered in the center of explosion. As a result, a shock with an almost spherical shape is reflected backward with the same velocity. The simulation also shows some departures from a spherically symmetric reflected shock especially when the shock is close to the center of the remnant. These departures are numerical artifacts introduced by the Cartesian mesh and, in fact, they increase as the shock focuses on the center of the remnant, namely following the decrease in the number of grid points per shock radius.}

{In the light of the 3-D MHD simulation discussed above, we conclude that the effect of RS bounce back \textit{from the center} is also present in 3-D simulations
if a spherically symmetric blast wave is assumed to propagate through a medium with spherically symmetric configuration. Therefore, we conclude that the RS bounce from the center is due to the symmetry introduced in the simulations and it
is neither a numerical artifact nor the effect of the reduced dimensionality. In fact, by 1-D runs, a perfect spherical symmetry is introduced in the system which ideally focuses the RS on the geometric center of the remnant (which is coincident with the center of the explosion). This effect is present even in 3-D simulations if they assume a spherically expanding blast wave (we have seen this in Animation \ref{animA3}). Conversely, if some asymmetry is introduced (for instance, because of a highly non-uniform environment or because of large-scale asymmetries in the explosion) the RS bounce would be very limited or it might occur in a point different from the geometric center. For instance, in a recent paper on the modeling of IC 443, the highly inhomogeneous ambient environment forces the RS to partially focus in a point which is largely offset from the center of explosion \citet[][see Fig. 2 and respective movie there]{2020arXiv201208017U}.
} {Likewise, motion of the progenitor star will induce an asymmetric wind bubble and hence incoherent reflection off the cavity wall \citep{2020MNRAS.493.3548M,2021MNRAS.502.5340M}.}

\section{Summary and conclusions}
\label{youngsnr:conclusions}

One-dimensional MHD simulations of SNR evolution have been performed for three models: SNIa, SNIc, SNIIP. Our simulations are magneto-hydrodynamic and enable us to study behavior of MF in young SNRs.
The conclusions of the present paper are as follows. 

\begin{enumerate}
    \item Early evolution of SNR with the power-law ejecta density profile follows the self-similar solutions by  \citet{1982ApJ...258..790C,1982ApJ...259..302C,1985Ap&SS.112..225N}. 
    \item All our three SN models evolve to Sedov stage once cooling is not essential. Distinctive feature of the Sedov phase is a constant ratio between kinetic and thermal energy. 
    \item Once radiative losses become effective, the adiabatic condition drops and SNR continue to evolve in the post-adiabatic regime. Then the shock is partially radiative. If the whole energy of a shocked fluid is quickly radiated away then SNR are in the fully radiative phase.
    \item Clear long-lasting Sedov stage in SNR evolution could rather be exception than the rule.
    \item The time of transition between the earliest ejecta-driven phase and the Sedov phase is rather long, with its distinctive physical features. It merits to be recognized as a separate phase in SNR evolutionary scheme.
    \item The main feature of this stage is energy conversion into different components.
    \item The updated scheme of SNR evolution could be a sequence of the following phases: free expansion (of gas) -- energy-conversion -- Sedov-Taylor -- post-adiabatic -- radiative.
    \item MF is dynamically unimportant until the end of the Sedov phase.
    \item 1-D simulations show no physical reason for the radial MF component to be dominant over the tangential one in young SNRs, if they were of the similar strength initially. 
    \item Radial MF experiences rapid drop right downstream of FS and we show analytically the reason for this effect.
    \item The polarization planes in the radio maps of such remnants have to be perpendicular to the radii contrary to the radial directions observed in a number of young SNRs.
    \item In order to explain the observed radially-aligned polarization patterns in a number of young SNRs, one needs to look for the effect in the 3-D MHD simulations which are able to catch the instabilities as a possible reason for such a polarization effect. 
\end{enumerate}

\section*{Acknowledgements}

{We thank the anonymous referee for careful reading of the manuscript.} OP acknowledges support of Theoretical Astroparticle Physics group for short stays at DESY, Zeuthen. Most of simulations were executed on a cluster at DESY. A part of simulations were performed on resources given in the frame of the Italian CINECA class C projects HP10CKMKX1 and HP10CR7V42. OP and TK acknowledge support from the Ukrainian project 0118U004780. 

\section*{Data availability}

Data available on request.

\bibliographystyle{mnras}
\bibliography{youngSNRs} 

\appendix
\section[]{Animations}

\subsection{}\label{animA1}
Animation shows evolution of the profiles of density $\rho$, pressure $p$, flow velocity $v$, tangential MF strength $B_\perp$, radial MF strength $B_\|$ in models for SNIa (red lines), SNIc (blue lines), SNIIP (green lines).

\subsection{}\label{animA2}
The same as on A1 but for the normalized profiles: $\rho/\rho\rs{s}$, $p/p\rs{s}$ where $\rho\rs{s}$ and $p\rs{s}$ are the immediately post-shock values. Vertical axis is linear at the left plots and logarithmic at the right plots. Self-similar solutions are shown by the dashed lines: yellow lines represents \citet{1982ApJ...258..790C} solution for $n=11$, $s=2$, grey lines are for the \citet{1959sdmm.book.....S} solution.

\subsection{}\label{animA3}

{Animation shows the evolution of the spatial distribution of mass density (in log scale) for a 3-D MHD simulation that describes the spherically symmetric blast wave originating from a SN explosion that propagates into a wind-blown bubble surrounded by a thin dense, cool shell (cavity wall) with spherical symmetry, originating from the interaction between the wind of the progenitor star and the surrounding ISM.}

\bsp	
\label{lastpage}
\end{document}